\documentclass{agujournal2019}
\usepackage{url} 
\usepackage{lineno}
\usepackage{soul}
\usepackage{booktabs}
\usepackage{multirow}
\draftfalse

\begin{document}

%
%


\title{Global climate modeling of the Martian water cycle. I)~Improved parameterizations of physical processes}
%
%




\authors{J. Naar\affil{1}, F. Forget\affil{1}, A. Bierjon\affil{1} , M. Vals\affil{2}, E. Millour\affil{1}, M. Wolff\affil{1,3}, F. González-Galindo\affil{4}, B. Richard\affil{1}}

\affiliation{1}{Laboratoire de Météorologie Dynamique, IPSL, Sorbonne Université, ENS, PSL research university, Ecole Polytechnique, CNRS, Paris, France}
\affiliation{2}{Laboratoire Atmosphères, Milieux, Observations Spatiales,
CNRS/UVSQ/IPSL, Guyancourt, France}
\affiliation{3}{Space Science Institute, Boulder, CO 80301, USA}
\affiliation{4}{Instituto de Astrofísica de Andalucía, IAA-CSIC, Granada, Spain}




\correspondingauthor{Joseph Naar}{joseph.naar@lmd.ipsl.fr}




\begin{keypoints}
\item Temperature-dependent cloud microphysics is a key improvement regarding the representation of water ice clouds and supersaturation in our model
\item Latent heat of sublimation is of second-order importance for the present-day water cycle but is of first-order importance for paleoclimatic studies
\item The martian water cycle is extremely sensitive to the coupling of physical processes, which calls for advanced numerical tuning methods
\end{keypoints}

%
%

%
%

\begin{abstract}

We implement four new physical parameterizations relevant to the water cycle in the Mars PCM (previously Mars Laboratoire de Météorologie Dynamique GCM)~: the latent heat of ground ice sublimation, a temperature-dependent nucleation scheme in cloud microphysics, differentiated albedos for water frost and perennial ice and scavenging of water ice clouds by CO$_2$ particle formation. After a new tuning of the poorly known model parameters, we find that combining these parameterizations allows for an improved agreement of the modeled water cycle compared to Thermal Emission Spectrometer column-integrated vapor and ice retrievals. We show that latent heat cooling prevents the formation of surface liquid water at present and likely in the recent past on Mars.  Using two albedos corrects the overestimation of the northern sublimation event, a major defect of the model. Our study highlights that each parameterization has indirect consequences amplified by the radiatively active clouds. We stress that the increasingly complex water cycle in the Mars PCM should be studied in the future using numerical methods tailored for climate modeling, in addition to the effect of each individual component.
\end{abstract}

\section*{Plain Language Summary}

Modeling the water cycle of the planet Mars sheds light on observations and data gathered from space missions. On Mars, the water cycle is driven by the sublimation of the Northern Polar Cap, which controls the amount of water in the Martian atmosphere. Despite the very low amount of water in the atmosphere, water ice clouds form and influence the overall behavior of the atmosphere. We have improved our physical description of ground ice deposits and water ice clouds. The incorporation of new, more realistic processes allows for a better understanding of space-based observations. Moreover, these processes are particularly significant in the modeling of recent paleoclimates of the planet Mars, during which ice clouds are much more abundant. Specifically, they suggest that the formation of liquid water at the surface is inhibited in the last several million years of Mars' history. The intricate coupling of all components related to the water cycle makes analyzing their impact on the climate a complex and challenging task. Future studies modeling the water cycle would benefit from employing dedicated numerical methods tailored for Global Climate Models analysis.

\section{Introduction}

Two major challenges compel us to model the martian global water cycle. First, we seek to explain the meteorological activity recorded by the available observations to further characterize of the martian climate and improve our understanding of atmospheric physics. Second, we need robust and trustworthy numerical climate models to investigate the evolution of the water cycle during Mars' past climates.  In particular, variations of obliquity and orbitally-driven forcing is thought to have had extensive consequences on the behavior and intensity of the global water cycle \cite{Habe:03,Misc:03}.

As the complexity of Global Climate Models (GCMs) increases, more and more features of the global water cycle can be investigated. Specifically, the implementation of Radiatively Active Clouds (RACs) by  \citeA{Made:12rad} in the Mars Planetary Climate Model (PCM, previously LMD-GCM) was a key improvement in the ability of the model to accurately compute the thermal structure of the atmosphere. Due to the strong coupling of water ice RACs with other components of the climate system, the simplified cloud scheme failed to reproduce the seasonal behavior of the water cycle. \citeA{Nava:14jgr} presented the implementation of a detailed water ice cloud microphysics scheme that includes ice nucleation on dust, particle growth and scavenging of dust induced by the condensation of ice. In that study, the global features of the water cycle were numerically constrained (i.e., tuned) using a total of 4 physical parameters (see table \ref{table:watparams}). These parameters control the properties of the surface over the Northern Polar Cap, the main atmospheric source of water vapor through summer sublimation, and the formation and evolution of water ice clouds. The restricted number of parameters allowed for a good physical interpretation, albeit it remains somewhat simplistic. Nevertheless, the simulated global water cycle by the model was in good agreement with observations from Mars Global Surveyor Thermal Emission Spectrometer (TES) \cite{Smit:01tes,Smit:02}.  However, the model still had shortcomings with respect to some of the observed features of the global water cycle. In particular, an overestimated sublimation peak at northern summer solstice, followed by a dry bias in the northern tropics during the second part of the year. The high opacity of clouds in the polar night was also inconsistent with observations. 

In this paper, we present the updated physical component of the Mars PCM. We have improved our model by adding physical processes to our description of ground ice reservoirs, cloud microphysics and aerosol interactions. We discuss the implementations and validate them against observations. We show that our modeled global water cycle of Mars is in significantly better agreement with the observations than the previous version.

Section \ref{section:model} summarizes the baseline model as well as the new physical improvements. Next, we present our new reference simulation and compare it with observations in section \ref{section:refsimu}. We discuss the individual effects of the latent heat of ground-ice sublimation and the implications for the modeling of surface liquid water in recent past climates in section \ref{section:lh}. We characterize the effects of the previously neglected temperature-dependent water ice cloud microphysics in section \ref{section:tempmsensi}. This is followed by a presentation of the basic parameterization for differentiated ground ice albedos in section \ref{section:albedo}. We discuss the effect of scavenging water ice particles with CO$_2$ ice crystals in section \ref{section:scavenging}. Finally, we present our conclusions and prospects in section \ref{section:ccl}.

\section{Model description}
\label{section:model}

The study is performed using the Mars PCM \cite{Forg:99,Forg:11}. The model consists of a combination of a generic dynamical core and a set of physical parameterizations specific to Mars. The grid-point ``LMDZ'' dynamical core solves the fundamental equations of fluid hydrodynamics on a standard 64$\times$48 longitude-latitude grid ($\sim$330$\times$220 km at the equator) with a Fourier filter applied at high latitude grid cells to account for the Courant-Friedrichs-Lewy criterion, and with hybrid $\sigma$-pressure vertical coordinates. The physical parameterizations (also known as the ``physics package'') are computed separately, column-wise in each grid cell of the model. Each cell also has a set of subsurface layers that include a temporal-spatial thermal inertia to compute the heat flux within the soil.

The CO$_2$ seasonal cycle is tuned using the surface CO$_2$ ice albedo and the subsurface thermal inertia so that the PCM reproduces the measured surface pressure variations associated with atmospheric condensation \cite{Hour:93}. The atmospheric dust aerosol spatial distribution is driven by atmospheric dynamics, but it is normalized to match the measured column optical depth daily scenario from \citeA{Mont:15,Mont:20}.  This is done using a ``semi-interactive'' two-moment scheme \cite{Made:11dust}. Airborne dust acts as condensation nuclei for water ice in the atmosphere, and thus the dust cycle is intrinsically linked with the water cycle. 

\subsection{Tuning the Global Water Cycle}
\label{subsec:gwatcycle}

The physical parameterizations for the Global Water Cycle used in this study are detailed in \citeA{Nava:14jgr}. Here, we summarize the key elements and refer the interested reader to this publication for a detailed description. 

Radiatively active clouds (RACs) are needed in GCM studies to reproduce the local thermal structure of the Martian atmosphere \cite{Made:12rad}. However, when it was implemented, the RAC-induced feedbacks strongly degraded the simulated water cycle, because of the radiative effects of polar summer clouds that are not observed in reality. \citeA{Nava:14jgr} addressed this issue by implementing an explicit cloud microphysics scheme. The scheme includes nucleation on airborne dust, supersaturation, ice crystal growth and sublimation, and scavenging of dust. 

The modeling of the climate of Mars corresponds to that of a chaotic system, which is particularly sensitive to the effect of RACs. Varying any tunable parameters of the water cycle influences the trajectory of the system. The primary tunable parameters are the effective variance of the log-normal distribution of cloud particle size $\nu_{\mbox {\scriptsize eff}}$ that affects the sedimentation of ice, and the contact parameter \textbf{m} which controls the fraction of activated nuclei. The corresponding baseline settings for these two parameters were $\nu_{\mbox {\scriptsize eff}}$~=~0.1 and \textbf{m}~=~0.95.

The atmospheric water content on Mars is dictated by surface exchanges with ground sources and sinks of water. Surface thermophysical properties, albedo and thermal inertia, are prescribed in the PCM using TES observations \cite{Putz:05,Wils:07}, typically ranging from 50 to 2000 J m$^{2}$ K$^{-1}$ s$^{-1/2}$ (tiu) for thermal inertia and from 0.3 to 0.5 for water ice albedo. Northern perennial surface water reservoirs are identified using TES surface temperatures. The corresponding grid cells are identified in the PCM in accordance with the latitudinal distribution of perennial ice identified in TES surface temperature data \cite{Nava:14jgr}. Thermal inertia and albedo are treated as tunable parameters in these grid cells, where we have adopted the reference values of 800 tiu and 0.35, respectively.

Assessing the individual effect of each of the main parameters summarized in table \ref{table:watparams} is complicated, because a subtle change in water ice cloud content or vertical structure can influence the total atmospheric water inventory. The method we chose for tuning the water cycle is to explore the most physically accurate parameter settings possible individually, and to assess their influence on well-documented water cycle features that are diagnostic because they can be interpreted physically and linked to general behaviors of the PCM. Specifically, the Northern sublimation event atop the Norther Polar Cap (NPC) around $L_S$=120$^\circ$, the water vapor integrated column in the tropics and the opacity of the Aphelion Cloud Belt (ACB) were identified as key structures to investigate. 

Tuning the parameters of the first martian GCMs was not too difficult because the different components of the water cycle (water vapour, clouds) had no effect on temperatures and the atmospheric circulation. However, the implementation of Radiatively Active Clouds in Mars GCMs initiated a paradigm change \cite{Made:12rad,Nava:14jgr,Kahre:17mamo}. Because of the intricate coupling with every component of the climate system, the investigation of the global water cycle is shifting from the incremental improvement of physical processes in the PCM to studying the coupling and emerging properties of all the components related to the water cycle \cite{Forg:14,Kahre:17mamo,Forg:22}. In this study, we thus implement 4 new physical enhancements and tune the PCM accordingly. We discuss their influence over the climate independently, in specific numerical experiments, to validate our understanding of each process. 

\begin{table}
\caption{Former tunable parameters of the Global Water Cycle used in \citeA{Nava:14jgr}. Thermal inertia unit (tiu) is J m$^{2}$ K$^{-1}$ s$^{-1/2}$.}
\centering
\begin{tabular}{l c}
\hline
  Parameter  & Reference value  \\
 \hline
   Contact parameter \textbf{m}  & 0.95   \\
   Effective variance for sedimentation $\nu_{\mbox {\scriptsize eff,sed}}$  & 0.1   \\
   Albedo of frost and permanent ice  & 0.35   \\
   Thermal inertia of permanent ice  & 800 tiu   \\
 \hline
\end{tabular}
\label{table:watparams}
\end{table}

\subsection{Time stepping of the model as a tuning parameter}
\label{subsec:timestep}

For convenience, 1 minute in the PCM is a $\frac{1}{24\times60}$ fraction of a Martian day, or \textit{sol}, set to 88775 seconds long. The standard Mars PCM uses two timesteps for integration, one each for the dynamics and the physics. They were set to 1.5 and 15 minutes, respectively. The dynamical timestep is disregarded for the rest of this paper: ``timestep'' refers only to the physical timestep in the model. During the early phase of this study, we found that decreasing the physical timestep of integration to 7.5 minutes significantly affected the behavior of the global water cycle. 

This time-stepping sensitivity is inherent to GCMs for two main reasons. First, the numerical methods used to solve the partial differential equations incorporate truncation errors which are affected by the integration step. Second, some processes have characteristic timescales smaller than the time-stepping of GCMs and are thus parametrized. The parameterizations are validated and then tuned for the model to match the target data. To some extent, the accuracy of the model regarding observations is partly due to compensating errors, including the time-dependent truncation previously mentioned.


The microphysics scheme of water ice clouds computes nucleation and crystal growth which may have timescales below the minute, depending on the environmental conditions. Due to the coupling of these non-linear rapid processes, a dedicated sub-timestep of 30 seconds was needed in \cite{Nava:14jgr} to prevent the unrealistic formation of clouds over the polar cap during Northern summer. 

The last reference setup of the model was performed with a specific timestep of 15 minutes. We now use a shorter, more realistic timestep of 7.5 minutes. The microphysics sub-timestep is also reduced to 15 seconds. The tuning of the parameterizations of the global water cycle must be adjusted to this particular configuration, regardless of the newly implemented physics. In the rest of this paper, time discretization is fixed and our parameterizations are evaluated in this specific framework. Time discretization in the PCM remains a major issue which is the dedicated subject of a companion paper to the present one.



\subsection{Physical improvements}
\label{subsec:physics}
\subsubsection{Latent heat of ground water ice sublimation}
\label{subsubsec:lh}

The latent heat flux associated with ground ice condensation and sublimation is usually neglected in PCM studies of the Martian global water cycle because of the low temperatures of water ice frost on Mars, corresponding to small energy fluxes \cite<e.g.>{Rich:02nat, Habe:19}. However punctuated events, such as the excavation of mid-latitude underground ice by impact craters, may expose large quantities of water ice to a relatively warmer environment. In addition, taking into account the latent heat flux is important when investigating past climates on Mars, because the cooling associated with summertime sublimation would strongly affect the temperature of mid-latitude ground ice deposits and their long-term fate \cite<e.g.>{Mont:07chap,Made:14ldm}.

We implement the following formulation of the latent heat flux for ground ice sublimation/condensation \cite{Mari:04}~:
\begin{equation}
    L_v = E_w \times 1000 \times (2834.3 - 0.28 \times (T_s -273.15) - 0.004 \times (T_s - 273.15)^2 )
\end{equation}
with $T_s$ the surface temperature [K] and $E_w$ the mass flux [kg/s]. This formulation is similar to that within water ice clouds in the PCM.  However, it  requires a specific numerical scheme with a subtimestep in order to properly account for the coupling between temperature and sublimation within the PCM physical timestep.

We have performed a numerical validation experiment with the 1D version of the PCM in order to assess the influence of latent heat on surface temperature and associated ice sublimation rates. The model is run for 100 sols before analyzing any results to ensure subsurface and boundary layer have reached steady-state conditions. The simulations are conducted with and without latent heat release under conditions observed by the Phoenix lander where water ice has been found right beneath the surface \cite{Cull:10}. That is assuming a background humidity of 10 precipitable microns, a surface wind of 10 m/s, albedo of 0.35, thermal inertia of 300 tiu and surface roughness of 1 cm, at a fixed solar longitude $L_S$=90$^\circ$ corresponding to the northern summer solstice. This approach was used in the martian NASA Ames GCM \cite{Mont:04phoenix}, with very similar results to ours.

\begin{figure}[h!]
\includegraphics[width=\textwidth]{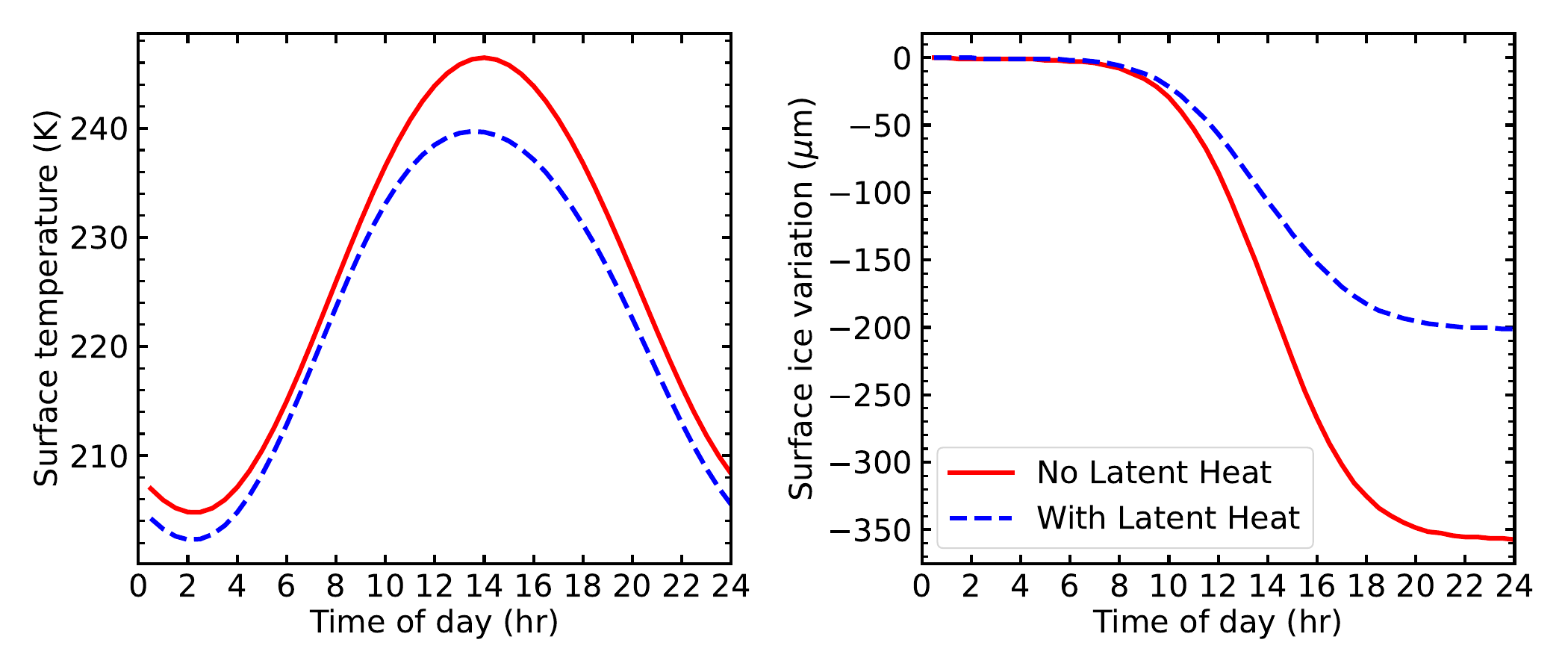}
\caption{Effect of latent heat of sublimation for exposed pure ice at 70$^\circ$N at summer solstice regarding surface temperature (left) and ice thickness variation (right).}
\label{fig:latent1d_base}
\end{figure}

Figure \ref{fig:latent1d_base} shows that in this experiment, latent heat release reduces diurnal heating of the martian surface by $\sim$ 8~K, with a subsequent decrease in the diurnal ice surface variation of about 40$\%$. The influence of the latent heat release on the global water cycle is more subtle in the PCM, as the water ice reservoirs are located at higher latitudes. This is discussed further in section \ref{section:lh}.

\subsubsection{Temperature-dependent water contact parameter}
\label{subsubsec:m}

\begin{figure}[h!]
\includegraphics[width=\textwidth]{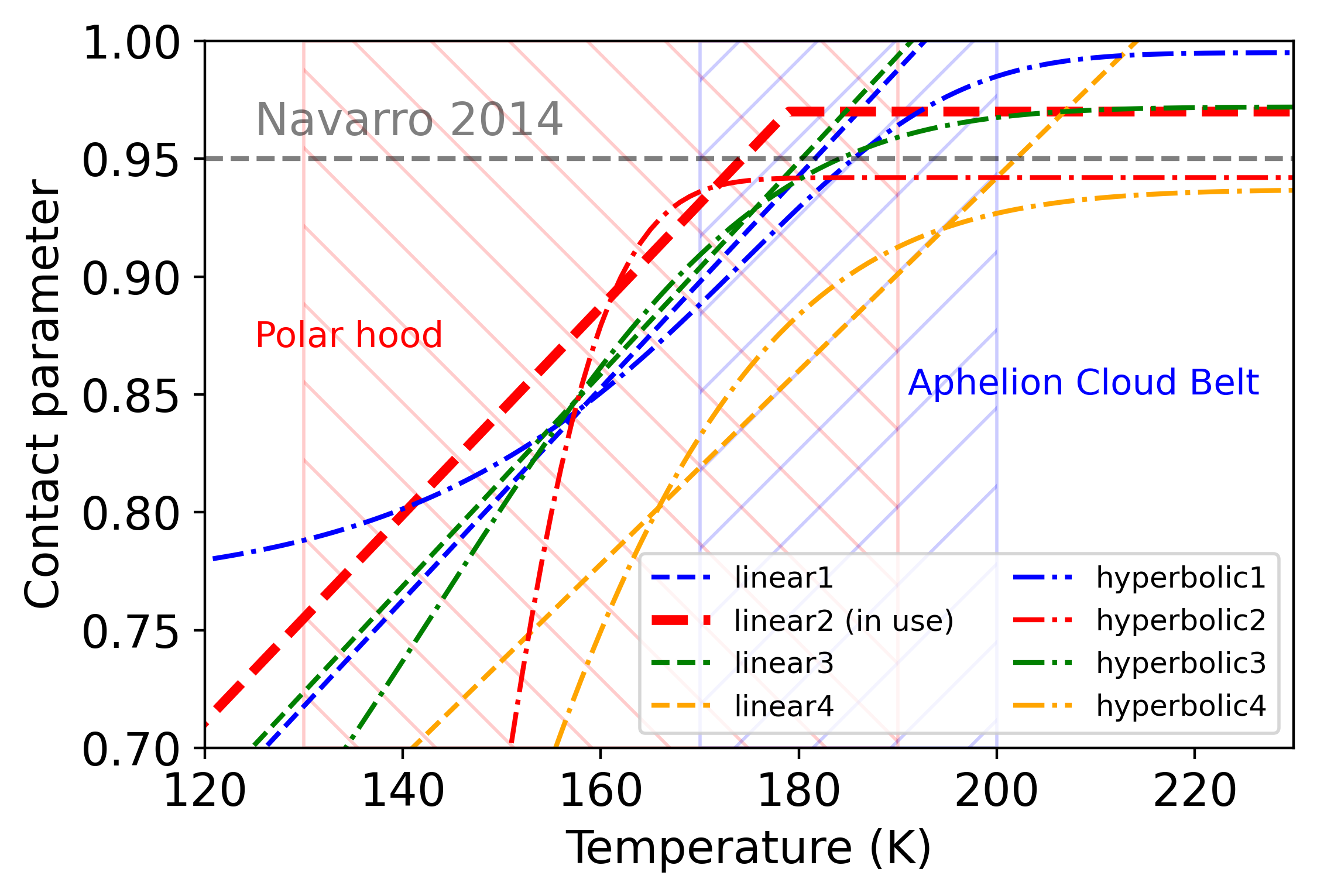}
\caption{The set of tested temperature-dependence parameterization for the water contact parameter from \citeA{Maat:14}. The water contact parameter is maxed out at 0.97 in the PCM for physical realism. The constant value of 0.95 used in \citeA{Nava:14jgr} is also displayed as a dashed grey line.}
\label{fig:temp_dep_m}
\end{figure}

The cloud microphysics scheme is based on the heterogeneous nucleation theory. This parameterization relies on the ``water contact parameter'' (m), which controls the fraction of airborne dust nuclei available for water condensation using a simplified spherical geometry approach \cite{Mont:02}. The water contact parameter cannot be easily derived from theory as it depends on the martian dust physical properties that vary over time, space and environmental conditions. Martian GCMs typically use contact parameters ranging from $\sim$ 0.95 to 0.975 \cite<e.g.>{Cola:99, Kahre:17mamo}, as in our reference simulation. However, theoretical and experimental work suggest a strong temperature dependence of the contact parameter, showing a dichotomy between high ($>170$ K) and low temperatures \cite{Trai:09,Irac:10,Pheb:11,Maat:14}. \citeA{Nava:14jgr} highlighted the influence of the contact parameter on the global water cycle in the PCM. Increasing the contact parameter tends to lower ice particle size for a given water-ice cloud mass as nucleation is favored over particle growth. This enhances the transport of polar water towards the tropics by the clouds. However, the increase of the contact parameter also amplifies the opacity of the polar hood during the second part of the year. Most of the time, our simulations showed a higher than expected atmospheric water sublimation peak in the polar region, and a thicker than expected polar hood. A temperature-dependent parameterization of the contact parameter addresses this specific issue. We have tested all the relations developed in \citeA{Maat:14}, shown in figure \ref{fig:temp_dep_m}, and find that the following formulation best reproduces the observed cloud opacity and water vapor column of TES~:

\begin{equation}
     m = min(0.0044*T + 0.1831,0.97)
\label{eq:tempdepm}
\end{equation}

This temperature dependence is linear, despite the authors stating that a hyperbolic relation is more accurate to describe the bimodal behavior of water contact parameter on Mars. There have been previous unsuccessful tests with implementing temperature-dependent contact parameter from \citeA{Trai:09} in the previous PCM release~: the lower nucleation rates would strongly inhibit high-latitude cloud formation, leading to colder polar regions and eventually drying out the water cycle. To the best of our knowledge, other Mars GCMs encounter the same issues and use a constant contact parameter \cite<e.g.>{Kahre:23}. The contact parameter relation we use (equation \ref{eq:tempdepm}) is the only one we tested with a high enough contact parameter in the Aphelion Cloud Belt temperature range (around $\sim$170~K, see figure \ref{fig:temp_dep_m}) to sustain cloud opacity comparable to observations. We thus use this simple linear relation and discuss the effect on the global water cycle in section \ref{section:tempmsensi}.

\subsubsection{Perennial ice and seasonal frost albedo discrepancy}
\label{subsub:alb}

The seasonal behavior of the northern polar cap is characterized by an evolution of surface water ice albedo through spring and summer \cite{Kief:01,Hale:05,Appe:11}. This behavior exhibits variations in space and time, depending on the region monitored and on the Martian year considered. The cause for these variations may be multiple, such as dust deposition, water ice grain size evolution or volatile condensation on cold trap slopes. The outlier perennial ice deposits seems to have a distinctly different albedo trend from the main cap, suggesting a potentially different subsurface ice inventory and thermal inertia. Nonetheless, the current treatement of the Northern polar cap is not realistic enough in the model to reproduce these observations. This is due to albedo and thermal inertia being prescribed as constant over the NPC. As a consequence, one well-identified bias in the PCM is the overestimation of the atmospheric water vapor column during the sublimation event of the NPC \cite{Nava:14jgr}. To address this issue, we have introduced a parameterization that distinguishes between perennial ice and seasonal frost.

\begin{figure}[h!]
\includegraphics[width=\textwidth]{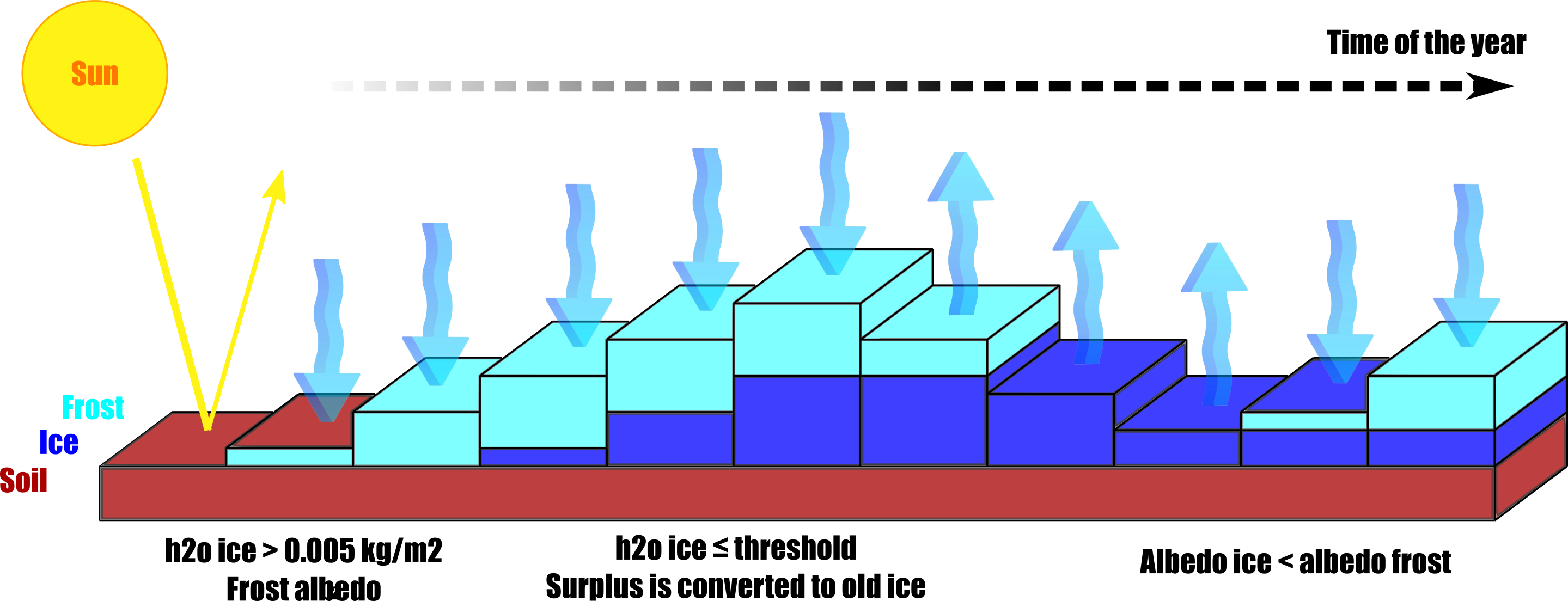}
\caption{Schematic illustration of our new parametrisation with differenciated perennial ice and frost albedos. }
\label{fig:schemaalb}
\end{figure}

This parameterization is illustrated in figure \ref{fig:schemaalb}. As of now, the only difference between seasonal frost and perennial ice is their albedo. In the PCM, frost albedo is active whenever frost thickness is above a threshold set to 5 millimeters (equivalent to 0.005 kg.m$^{-2}$). If the frost layer is thinner, the albedo is set to the standard value of the mesh bare ground derived from TES observations, unless the mesh is a perennial ice reservoir whose albedo is thus a tunable parameter. When frost accumulates on perennial ice over the NPC, we limit its growth up to a maximum threshold. If frost is deposited beyond this threshold, the surplus will be converted to perennial ice as to represent ice metamorphism and subsequent albedo evolution. By doing so, we ensure that seasonal sublimation will occur even if frost was massively deposited in the same grid cell during the previous winter. 

The determination of the two albedos for seasonal frost and perennial ice, as well as the threshold for maximum frost deposition, is detailed in section  \ref{section:albedo}. 

\subsubsection{Water ice scavenging by CO$_2$ ice}
\label{subsubsec:param_scavenging}

CO$_2$ is the main component of the Martian atmosphere. In cold regions like the polar nights and the mesosphere, the temperature can decrease enough to reach the condensation point of CO$_2$, forming frost at the surface as well as ice clouds in the atmosphere. Atmospheric condensation/sublimation of CO$_2$ is handled well in the model and accurately reproduces the seasonal pressure cycle measured by the different landers \cite{Forg:98}.
However, unlike the dust scavenging by water ice particles already implemented in the PCM \cite{Nava:14jgr}, there was no representation of the aerosol scavenging by CO$_2$ snow.

We introduce a parameterization to describe how CO$_2$ condensation affects the distribution of dust and water ice particles, which are used as condensation nuclei for CO$_2$ \cite{Maat:05}. In practice, the scavenging of aerosols can happen either by nucleation or by collision, where the latter occurs when the CO$_2$ ice falls and drags other particles. In our model, when the CO$_2$ condensation occurs in the whole column -- and not only in a cloud at a given altitude -- we consider that the effect of collisions can be treated as scavenging by nucleation. We adapt a common parameterization from Earth modeling \cite<e.g>{Jous:90} by assuming the aerosol mixing ratio within CO$_2$ snow is proportional to the local aerosol concentration. We thus use the scavenging ratio as a tuning parameter~:

\begin{equation} 
{\cal R}_{{\mbox {\scriptsize scav}}}  =  \frac{q_{{\mbox {\scriptsize aerosol\ in\ CO$_2$\ snow}}} }
                  {q_{{\mbox {\scriptsize aerosol\ in\ air}}} }
\label{eq:scav}
\end{equation}
with $q$ the mass mixing ratio (MMR) in either air or CO$_2$ snow [kg$_{{\mbox {\scriptsize aerosol}}}$/kg$_{{\mbox {\scriptsize environment}}}$].\newline

In this work, we assume a constant value for ${\cal R}_{{\mbox {\scriptsize scav}}}$. We employ the scheme from \citeA{Forg:98}, which computes CO$_2$ condensation/sublimation in each layer over the full atmospheric column and updates the CO$_2$ budget directly on the ground, starting from the highest layer $l=N$ down to the layer just above the surface $l=1$. The CO$_2$ ice mass flux $F\mathrm{ice}$ which sediments from layer $l+1$ into layer $l$ is then~:

\begin{equation} 
F\mathrm{ice}_{l+1}= 
\sum_{k=l+1}^{N} \frac{\partial m_k}{\partial t}
\end{equation} 

with $l$ a given atmospheric layer, $M_l$ its mass [kg$_{{\mbox {\scriptsize air}}}$.m$^{-2}$], $m_l$ the CO$_2$ ice mass and $\partial m_l / \partial t$ [kg$_{{\mbox {\scriptsize ice}}}$.m$^{-2}$.s$^{-1}$] the CO$_2$ condensation/sublimation rate.

In the case of condensation, that is if $\partial m_l / \partial t > 0$ (CO$_2$ ice forms in the layer), equation \ref{eq:scav} becomes~:
\begin{equation} 
{\cal R}_{{\mbox {\scriptsize scav}}} =\frac
      { 
            {- \left( \frac{\partial q_l}{\partial t} \right)_{\mathrm{scav}} }}
      {
            {\left( \frac{1}{M_l} \right) \left( \frac{\partial m_l}{\partial t} \right) q_l} }
\label{eq:cond}
\end{equation} 
 
Equation \ref{eq:cond} can be integrated over the physical timestep to obtain the mass mixing ratio of aerosol scavenged $\delta q_l$ from a given layer~:

\begin{equation}
\left( \frac{\delta q_l}{\delta t} \right)_{\mathrm{scav}}
= -\frac{q_l}{\delta t} \: \left( 1 - \mbox{e}^{{
      - \left( \frac{{\cal R}_{{\mbox {\scriptsize scav}}}}{M_l} \right) \left(\frac{\partial m_l}{\partial t}\right) \delta t}
            } \: \right)
\end{equation}

This implicit numerical scheme prevents the model from scavenging more than the initial aerosol mass in the layer.

In the case of sublimation, that is if $\partial m_l / \partial t < 0$ (CO$_2$ falling ice sublimates inside a given layer if it is warmer than the CO$_2$ condensation temperature), we assume that the release of the nuclei from the CO$_2$ snow is proportional to the sublimation rate~:

\begin{equation}
\left( \frac{\delta q_l}{\delta t} \right)_{\mathrm{scav}} =
  -\frac{\frac{\partial m_l}{\partial t}}{M_l} \: 
\frac{ F\mathrm{aer}_{l+1}} { F\mathrm{ice}_{l+1}}
\end{equation}

with $F\mathrm{aer}_{l+1}$ the aerosol flux scavenged from layer $l+1$ into layer $l$:
\begin{equation} 
F\mathrm{aer}_{l+1}= 
\sum_{k=l+1}^{N} \frac{\delta q_k}{\delta t}
\end{equation} \newline

The scavenging rate for the number of particles $N_l$ [1/kg$_{{\mbox {\scriptsize air}}}$] is derived in each layer from the mass tendency $\delta q_l/\delta t$, via the relation~:  

\begin{equation}
     N_l =
    \frac{3}{4}\frac{q_l}{\pi \rho_{{\mbox {\scriptsize aer}}}}
    \left( \frac{1+\nu _{{\mbox {\scriptsize eff}}}}{r_{{\mbox {\scriptsize eff}}_l}}\right) ^3
\end{equation}

with $r_{{\mbox {\scriptsize eff}}_l}$ the effective size computed by the PCM before the scavenging, and the fixed parameters $\nu_{{\mbox {\scriptsize eff}}}$ and $\rho_{{\mbox {\scriptsize aer}}}$ the effective size variance and the aerosol density [kg$_{{\mbox {\scriptsize aer}}}$.m$^{-3}$] respectively. For dust, we have $\nu_{{\mbox {\scriptsize eff}}}$=0.5 and $\rho_{{\mbox {\scriptsize dust}}}$=2500kg.m$^{-3}$ in the PCM. For water ice particles, it is three tracers that are being scavenged by CO$_2$ snow: the water ice mass $q_{H_2O ice}$, the water ice cloud condensation nuclei (CCN) mass $q_{ccn}$, and the number of particles, which is the same for water ice and CCN: $N_{H_2O ice}$=$N_{ccn}$. Since water ice CCN are purely made of dust in the PCM, we use the dust $\nu_{{\mbox {\scriptsize eff}}}$ and $\rho_{{\mbox {\scriptsize dust}}}$ properties to compute the tendency on $N_{ccn}$ from the one on $q_{ccn}$.

The scavenging ratio is therefore the only tunable parameter in this parameterization. In section \ref{section:scavenging}, we detail its influence on the polar night conditions, and how this process affects the global water cycle.


\section{Reference simulation and comparison with observations}
\label{section:refsimu}

\subsection{New reference Global Water Cycle}
\label{subsec:ref_wc}

All the simulations presented in this study are computed for Martian Year (MY) 26, because it has good TES data coverage \cite{Smit:04} and is representative of a standard Martian year without global dust storm. The PCM dust cycle is nudged in the simulations using the scheme from \citeA{Made:12clouds} and the Martian Year 26 scenario from \citeA{Mont:15}. The simulations are performed using the standard 64 $\times$ 48 PCM resolution with 26 vertical levels reaching a pressure of 0.2~Pa, corresponding to a pseudo-altitude of about 80 km. The results that are analyzed are taken after 10 years of model time computation to ensure that a steady-state is reached. Our results represent a new reference simulation for the Global Water Cycle, one that includes the latent heat of ground ice sublimation, a temperature-dependent contact parameter \textbf{m}, and a ground ice albedo of 0.33.

\begin{figure}[h!]
\hspace{-80pt}
\includegraphics[width=1.4\textwidth]{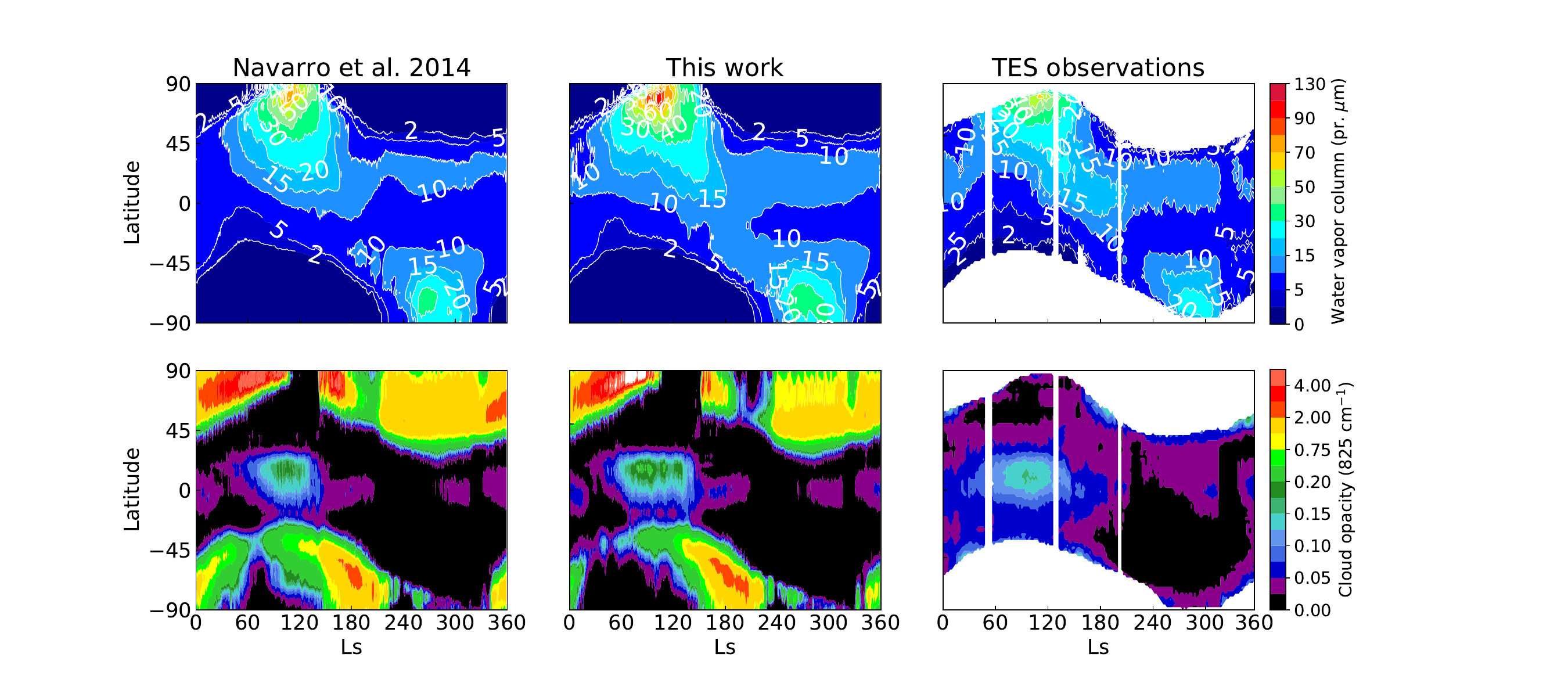}
\caption{Zonally averaged quantities at 2 pm local time for water vapor column in the different versions of the PCM and MY26 TES observations (top panels), and corresponding cloud opacity (bottom panels). Results are shown after 10 years of simulation.}
\label{fig:al033_wc}
\end{figure}

We present our reference simulation with latent heat of ground ice and the temperature-dependent contact parameter in figure \ref{fig:al033_wc} (middle panels). The differenciated albedos and the scavenging by CO$_2$ are implemented from this reference simulation and discussed in dedicated sections \ref{section:albedo} and \ref{section:scavenging}. We manage to obtain integrated water vapor columns and cloud opacity in agreement with TES observations.

The first region of interest is the Northern Polar Cap around the sublimation event. We see that in our new setup, because we slightly lowered the frost albedo from 0.35 to 0.33, the integrated water vapor column is overestimated compared to both the previous model version and TES observations. Tuning the albedo further down makes for a reasonable sublimation event at the cost of global atmospheric water inventory, which in turn degrades our global water cycle. 

The representation of tropical water vapor column during Northern fall is improved in our new reference simulation. Martian GCMs historically struggle to match TES observations as they have a dry bias at this time of year \cite{Nava:14jgr,Kahre:17agu,Habe:19}. This is linked to the meridional transport of water~: with our temperature-dependent contact parameter, the latitudinal transport is reduced and water may reside longer in the mid-latitude regions.

The temperature-dependent contact parameter also reduces the opacity of clouds in the polar nights. Specifically, there is now a striking absence of cloud in the northern polar hood around $L_S$=200$^\circ$, which is clearly visible in our new simulation (figure \ref{fig:al033_wc}, bottom third panel) and is somewhat comparable to Planetary Fourier Spectrometer (PFS) data \cite{Giur:21}.

The Aphelion Cloud Belt (ACB) has a higher opacity in our new simulation than in the previous PCM iteration. This is also linked to the temperature-dependent contact parameter and discussed in section \ref{subsec:mrsize}.

An additional notable feature is the southern sublimation event, whose amplitude is historically overestimated in the PCM. The model predicts its peak to occur around $L_S$ 255$^\circ$, $\sim$20$^\circ$ earlier than what is observed by TES, as can be slightly seen on figure \ref{fig:al033_wc} (and more clearly on figure \ref{fig:global_mtot_lh}). This offset was already identified in \citeA{Nava:14jgr} and is not noticeably affected by the water-related physical implementations we have tested. It is likely related to the seasonal timing of the CO$_2$ ice cap sublimation, as previously identified in \citeA{Nava:14jgr}, which is beyond of the scope of our study.

\subsection{Seasonal behavior of water in the PCM}
\label{subsec:groundice_ref}

\begin{figure}[h!]
\includegraphics[width=0.6\textwidth]{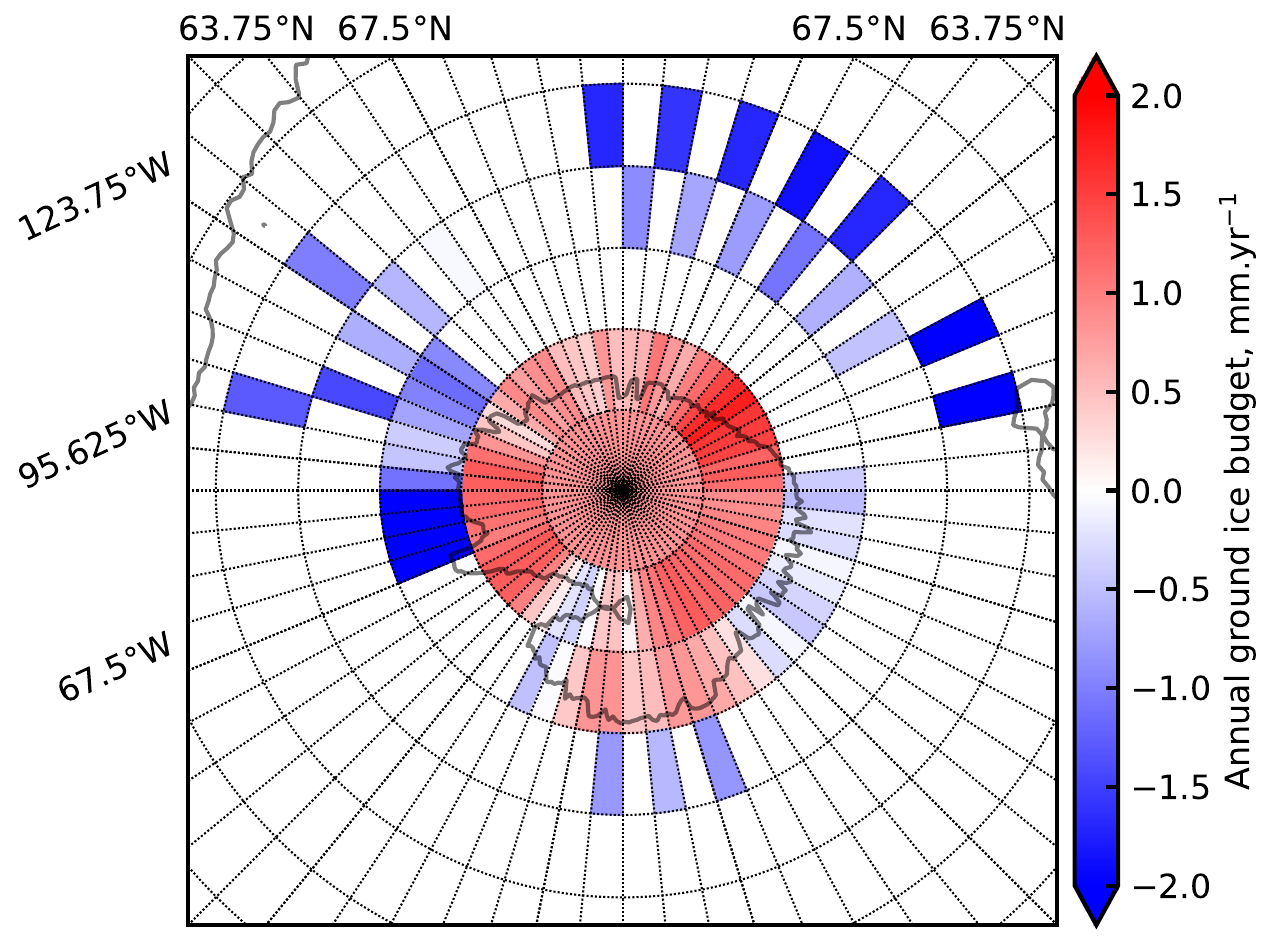}
\caption{Water flux of northern water cap}
\label{fig:watbudg}
\end{figure}

Figure \ref{fig:watbudg} shows the annual budget of ground ice atop the Northern polar cap. On average the total mass budget is a net loss of $2.7 10^{11}$ kg of ground water ice at the end of the year, that is an average of 0.2 kg.m$^{-2}$ sublimated ground ice on each polar cap grid cell. The actual amount of sublimated water is higher, but the central cap acts as a cold trap and a large fraction of the freshly available water vapor condenses right back on the NPC. High-resolution simulations with the Mars PCM have been performed and discussed by \citeA{Pott:17} in which the authors demonstrate that the spatial resolution employed here is insufficient to properly model the baroclinic wave activity at the cap edge. This wave activity is the main driver of the seasonal behavior of ice at the pole. As of now, we prescribe a constant albedo and thermal inertia over the whole NPC in the PCM.  However, since we could not find a more optimal set of parameters than the one used in this simulation (see section \ref{subsec:gwatcycle}), the northern water ice sublimation event remains overestimated in our reference simulation. We further address this topic in section \ref{section:albedo}.

\begin{figure}[h!]
\includegraphics[width=\textwidth]{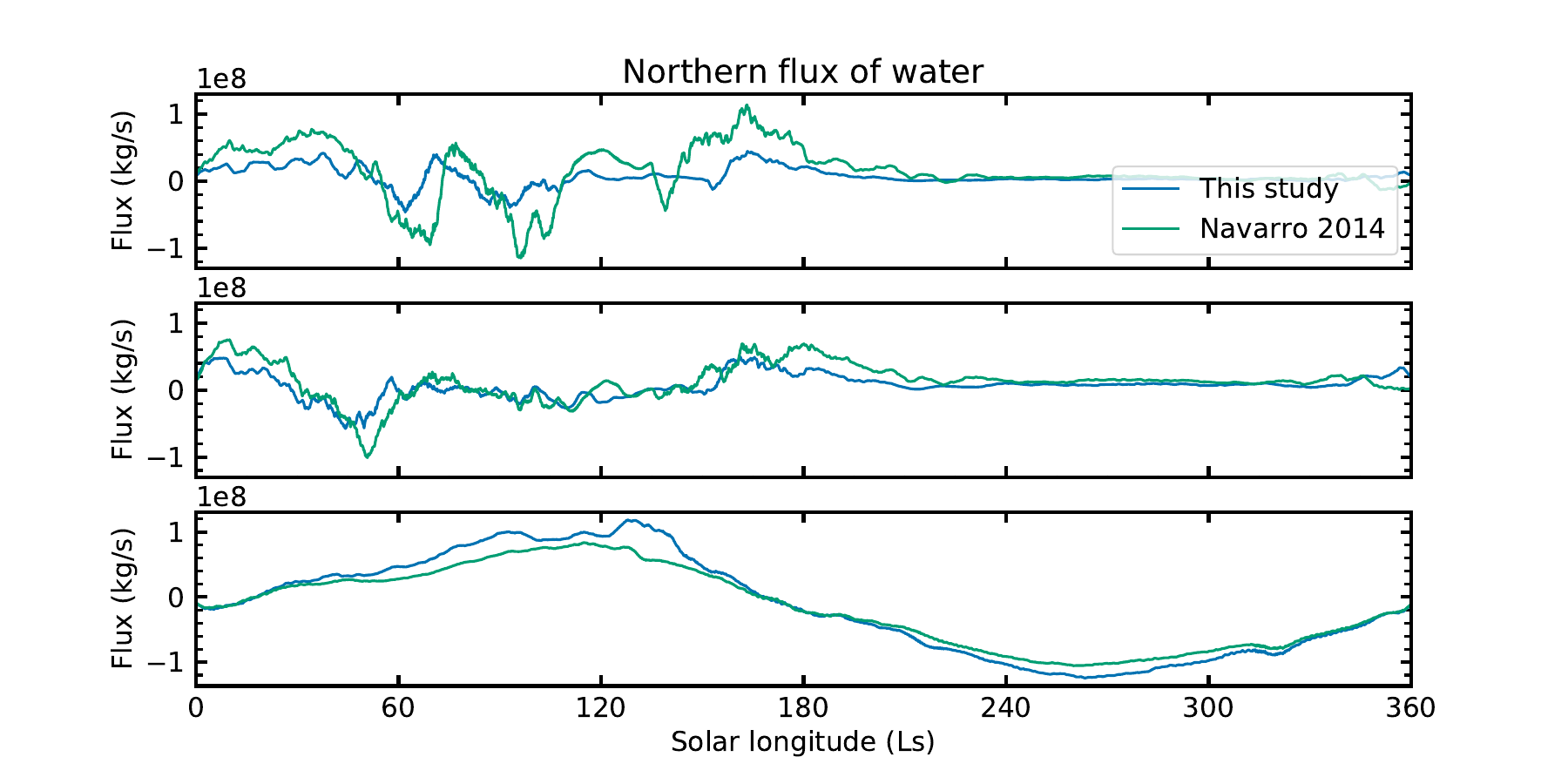}
\caption{Latitudinal transport of water through parallels 75$^\circ$N (up) and 67.5$^\circ$N (middle), and the equator (bottom). The flux is positive when southward.}
\label{fig:latflux}
\end{figure}

Following the northern sublimation peak, water vapor and ice are transported south towards the northern tropics and the southern hemisphere. This flux is evaluated in figure \ref{fig:latflux} with the following equation \cite{Pott:17}, along a given parallel~:

\begin{equation}
    \frac{\mathrm{d}m_{H_2O}}{\mathrm{d}t} =
    \int_{0}^{2\pi} \int_{0}^{z_{top}} \rho_{air}~q_{H_2O}~v~Rcos(\theta)~ \mathrm{d}\phi \mathrm{d}z
\label{eq:flux}
\end{equation}

with $\frac{\mathrm{d}m_{H_2O}}{\mathrm{d}t}$ the northward flux of water (positive toward north), $\rho_{air}$ atmospheric density, $q_{H_2O}$ the mass mixing ratio of water (ice and vapor), $v$ the meridional wind, $R$ the planetary radius, $\theta$ and $\phi$ the latitude and longitude, and $z$ the vertical level.

The upper panel shows the flux of water through the parallel located at 75$^\circ$N. Water is transported outside the polar region starting from around $L_S$$\sim$40$^\circ$ when the Northern frost starts subliming. Atmospheric water partly condenses outside the edges of the northern cap and is re-mobilized toward the pole around $L_S$$\sim$75$^\circ$ in the previous reference simulation as already noted in \citeA{Pott:17}, whereas in this study this effect is diminished by the new physical parameterizations. The middle panel shows the amount of water transported toward the mid-latitude regions. Overall, we find that our new reference simulation roughly reduces the amount of latitudinally transported water integrated across the northern polar regions by a factor of 2.

\subsection{Supersaturation of water vapor}
\label{subsec:msatu}

Water vapor is found in a supersaturated state throughout the atmosphere of Mars at all times of the year \cite{Malt:11}. This directly affects the global water cycle and its evolution through time as supersaturated vapor at the top of the atmosphere is photodissociated and escapes toward space \cite{Chas:04esc}.

\begin{figure}[h!]
\includegraphics[width=\textwidth]{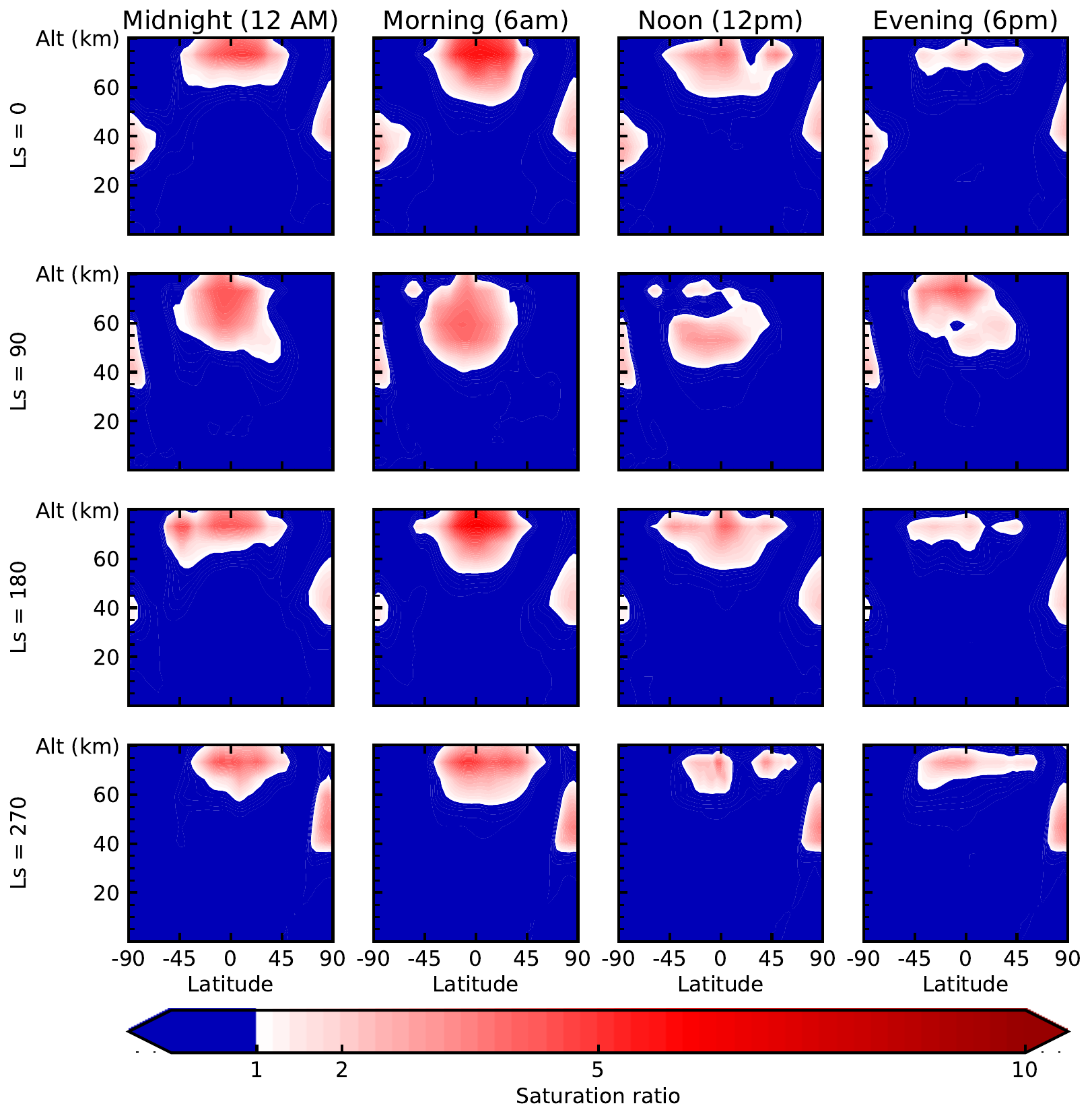}
\caption{10-day average of the zonal distribution of the saturation ratio at selected local times and solar longitude. A saturation ratio below 1 means the atmosphere is unsaturated (blue).}
\label{fig:saturation}
\end{figure}

Figure \ref{fig:saturation} shows the saturation ratio averaged over 10 days at 4 local times. We find that the altitude of supersaturated levels and their diurnal cycle matches the existing datasets \cite{Malt:11,Holm:22}. We note that the second cloud-free season in the Northern polar night at $L_S$=200$^\circ$ mentioned in section \ref{subsec:ref_wc} corresponds to a supersaturated atmosphere above the NPC. To the best of our knowledge, our simulation is the first to reproduce this feature consistent with both Mars Climate Sounder (MCS) and Planetary Fourier Spectrometer (PFS) data \cite{Bens:11,Giur:21}. Even though the decrease in cloud opacity was already visible in the previous reference state of the PCM (figure \ref{fig:al033_wc}), the temperature-dependent nucleation scheme now clearly prevents cloud formation in this region at this time of year.

Observations suggests that global dust activity drives dramatic changes in atmospheric escape rates of water, linked to supersaturated water vapor \cite{Chaf:14,Fedo:18,Stch:22}. This is frequently investigated with GCM studies \cite<e.g.,>{Vals:22}. The implementation of a temperature-dependent contact parameter is thus crucial in future studies.

\newpage
\section{Latent heat of ground water ice sublimation}
\label{section:lh}
\subsection{Sublimation of the Northern polar cap}
\label{subsection:lhsensi}

\begin{figure}[h!]
\includegraphics[width=\textwidth]{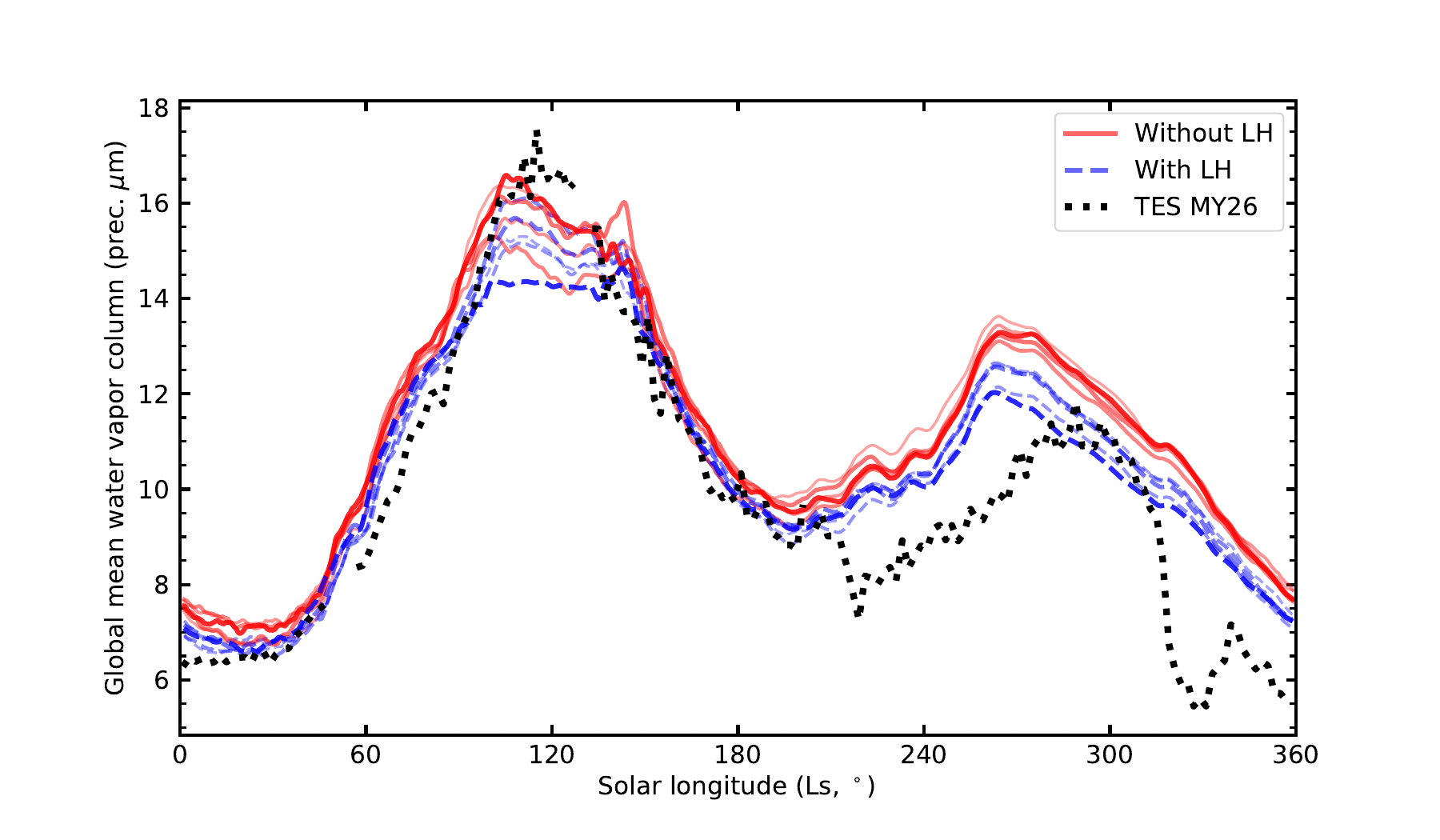}
\caption{10 year evolution of globally-averaged atmospheric water vapor for our experiment without latent heat of ground ice, or control experiment which consists in continuing the reference simulation for 10 years. Only the last 5 years are represented. The most recent years are thicker. We have also represented TES observations for MY26.}
\label{fig:global_mtot_lh}
\end{figure}

In order to assess the sensitivity of the water cycle to the latent heat of ground ice sublimation, we conduct 10-year simulations starting from the converged reference state presented in section \ref{section:refsimu}, with the latent heat turned off. However, distinguishing the direct effect of latent heat from the other water-related processes is challenging because of the strong feedback of the radiatively active water ice clouds, which amplifies any change in the water cycle. Figure \ref{fig:global_mtot_lh} shows the globally-averaged water vapor column for each of the last five years in the experimental simulation without latent heat and for the control experiment, compared to the TES measurements. When latent heat is neglected, the total amount of water vapor is higher by up to $\sim$1 precipitable micron in the experiment compared to our control simulation during the northern and southern sublimation periods. 

When looking at the instantaneous energy fluxes during summertime sublimation, we find that the latent heat accounts for less than 2~W.m$^{-2}$ integrated over the NPC throughout the whole sublimation season. However, this small difference in energy budget builds different trajectories over the passing years. 

Figure \ref{fig:global_mtot_lh} also illustrates the interannual variability of the model. The years are somewhat converged in the sense that global atmospheric water inventory is the same at the beginning of the year. But the contigent location of ground water ice deposits and subsequent consequences on soil temperatures may slightly vary. In addition, the water cycle budget is not actually closed (see section \ref{subsec:groundice_ref}). The timescale for simulations to reach a converged state is 10 years, but the model exhibits some interannual variability which needs to be quantified further in the future.

\subsection{Sublimation of mid-latitude excavated ice}
\label{subsection:1dlh}

The implementation of latent heat of groundwater ice sublimation has little influence on the water cycle within the PCM because of the high latitude of groundwater ice reservoirs and their low temperature. 
However, the energy flux related to latent heat of water ice sublimation becomes critical when modeling warmer ground water ice reservoirs. Such conditions are expected to be met both over the NPC and in the mid-latitudes during high obliquity excursions, which have happened periodically during the last few million years on Mars \cite{Head:03nat,Levr:04,Made:09}. Punctuated ice excavation events by meteorite impact can also expose bulk ice to atmospheric exchanges outside the polar regions. 
\citeA{Byrn:09} references several recent mid-latitude craters seen in HiRISE images that are consistent with excavated water ice. They suggest a protocol to assess the total amount of sublimated water ice within a crater, using a one-dimensional subsurface thermal model. They find an equivalent thickness of vaporized groundwater ice of a few milimiters, as a lower limit, for the first crater site studied (Site 1) during Martian Year 29 northern summer (between $L_S$=81$^\circ$ and $L_S$=111$^\circ$), at a latitude of 46.33$^\circ$N. \citeA{Dund:10} build upon this work and compute precise sublimation rates using a 1D layered thermal model, with reconstruction of environmental conditions happening at this site.

In order to assess the individual effect of latent heat cooling regarding ice sublimation rate in the PCM, we conduct a set of simple numerical experiments using the 1d version at the same time and location. The atmospheric water vapor content is fixed to 10 precipitable microns, a standard average martian value, assuming a well-mixed vertical profile. The albedo of bare ground is set to 0.25 (the standard albedo in use in the PCM at this location). Albedo of exposed frost and thermal inertia are fixed in all experiments. The investigated subliming ice at this Site 1 is in a flat crater. We find that a favorable set of parameters allows the surface temperature to rise above the melting point of water when ice is exposed on a 20$^\circ$ sun-facing slope, we thus set our base experiment in this configuration.

We run the model for several years to make sure the temperature profile in the soil is stabilized. Then, at $L_S$=91$^\circ$, we expose 10 millimeters of ice at the surface and monitor the sublimation rate and surface temperature cycle with and without latent heat cooling, for a range of albedo, wind velocities, and thermal inertia. The explored range for these parameters is summarized in table \ref{table:1dstudy}.

\begin{table}[h!]
\caption{Parameters range for the 1d latent heat cooling experiment. Reference study by \citeA{Dund:10} uses an albedo of 0.24 for bare ground and a thermal inertia of 233 J~m$^{-2}$~K$^{-1}$~s$^{-1/2}$.}
\centering
\begin{tabular}{l c}
\hline
  Parameter  & Considered values  \\
 \hline
   Latent heat release & .true. ; .false.   \\
   Frost albedo  & 0.3 ; 0.35 ; 0.4   \\
   Ground thermal inertia (tiu)  & 240 ; 400   \\
   Wind speed (m/s) & 2.5 ; 5 ; 10  \\
 \hline
\end{tabular}
\label{table:1dstudy}
\end{table}

\begin{figure}[h!]
\includegraphics[width=\textwidth]{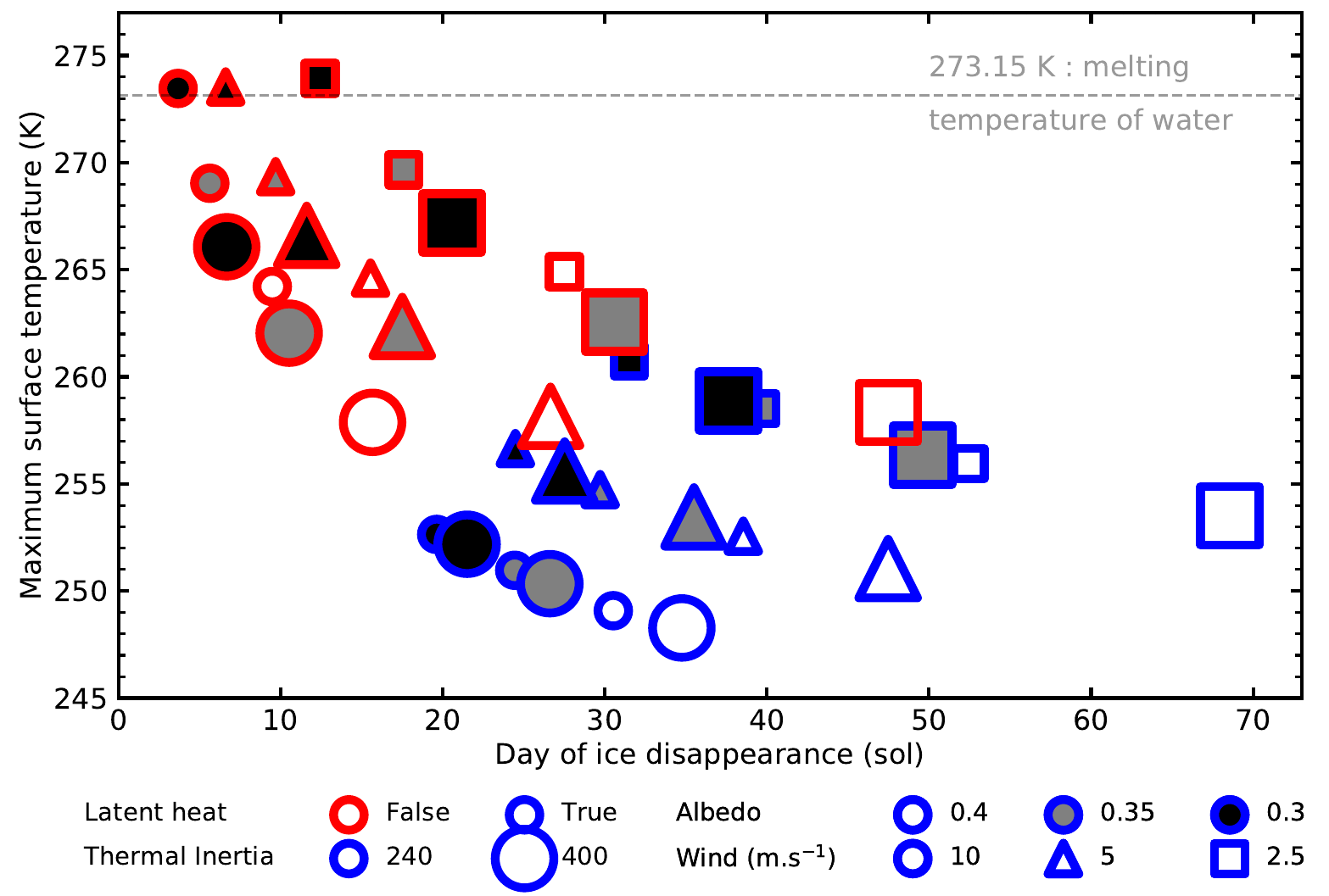}
\caption{Persistence of 10 mm exposed ground water ice on a sun-facing 20$^\circ$ slope at 46.3$^\circ$N latitude at $L_S$=91$^\circ$ using the 1d version of the PCM. Albedo of bare ground is set to 0.25 before ice is exposed at the surface. The experiment are run for a year before exposing the ice.}
\label{fig:diagram_tmax}
\end{figure}

Figure \ref{fig:diagram_tmax} shows the results of our 36 experiments with a varying set of parameters. The sublimation is computed with the surface turbulent flux \cite{Forg:99}, the sublimation rates are thus highly sensitive to wind speed. The free-convection regime should dominate at wind regimes below $\sim$2~m/s \cite{Hech:02}, which is not yet implemented in the Mars PCM. The exploration of this region of the parameter space is thus discarded at the moment. Albedo and thermal inertia control the surface temperature and thus the equilibrium saturation pressure of water above the surface, which also controls the sublimation. The latent heat cooling is stronger where the sublimation rate is maximal. We find that, regardless of the experiment, the latent heat cooling associated with sublimation increases the life expectancy of exposed ice by $\sim$20~sols. 

With our fixed water atmospheric profile, a frost albedo of 0.4 and a surface thermal inertia of 400 J~m$^{-2}$~K$^{-1}$~s$^{-1/2}$, we also have conducted an experiment using a flat surface, similar to \citeA{Dund:10} (not shown). We find that the 10 milimeters of ice are sublimated after 162 sols, which is about $L_S$=170$^\circ$. \citeA{Dund:10} find with their model a total of 6.9 millimeters in their most extreme scenario for the same duration. Their thermal regolith model include tabulated values of subsurface water ice from \cite{Byrn:09}, which dampens the thermal amplitude compared to our fixed thermal inertia. The representation of subsurface ice in the PCM is currently in development to further improve our representation of the soil in the model \cite{Lang:23ice}.

\subsection{Paleoclimatic studies and transient liquid water on Mars}

The possibility of liquid water on the surface of Mars is a major topic of interest, because its possible implications for soil chemistry and exobiology. Some of the geologically young or even present-day flow features observed, in particular gullies, may by formed in presence of liquid water \cite<e.g.,>{Mali:00,Cost:02,Harr:15}. Although the current paradigm has shifted in favor of dry processes \cite{Pilo:16} the scientific debate is still open.

GCMs compute surface pressure and soil temperatures at every time and location on the planet, and are thus a privileged tool to investigate the question of surface ice melting. Pioneer work on this matter by \citeA{Habe:01} identified the present-day martian regions where thin surface ice deposits could melt, that is where temperature is above the triple point of water but below the boiling point (with respect to local surface pressure and temperature). In the Northern Hemisphere, these regions are located between latitudes 0$^\circ$ and 30$^\circ$ where solar heating is strong enough, and in impact basins for the Southern Hemisphere. This work was extrapolated to the recent past of Mars in \citeA{Rich:05} by conducting numerical simulations at higher obliquities and with increased surface pressure, which favors ice melting conditions. Since then, \citeA{Made:14ldm} showed that the strong greenhouse effect of radiatively active clouds radically affects the modeled climate at high obliquity. \citeA{Dickson:23} conducted high obliquity simulations using the Mars PCM, and stated that liquid water might have occurred on bare ground locations where surface temperature exceeds 273~K. However, none of these GCM experiments account for surface water ice, and the subsequent latent heat warming for surface temperature computations.

With our set of 1d simulations, our maximal daily surface temperature reaches 285~K (in our experiments with TI=240 tiu, and 275~K with TI=400 tiu) even if we are at 46.33$^\circ$N, since we are on a 20$^\circ$ sun-facing slope. Then, the exhumated ice increases surface albedo. In our warmest simulations (without the latent heat cooling~: figure \ref{fig:diagram_tmax}, three top-leftmost simulations), surface temperature remains right above the melting point of water while ice is present, corresponding to Transient Liquid Water Potential (TLWP) conditions investigated in \citeA{Rich:05}. However, in our experiment, the associated latent cooling with the corresponding sublimation rates decreases the maximum surface temperature by at least $\sim$15~K, and up to $\sim$23~K in the fastest subliming case with the highest wind speed (figure \ref{fig:diagram_tmax}, largest red and blue circles). We could not find any 1d setup of the Mars PCM which allows for surface temperature to rise above the melting point of water when computing latent heat cooling associated with sublimation of ground ice.

To further expand our understanding of the possibility of transient liquid water on Mars through surface ice sublimation, GCM experiments must thus account for the presence of ice and the latent heat cooling associated. In addition, Martian slopes may cause local microclimate more favorable to the TLWP than the grid-mesh average: the statistical representation of slopes is currently being implemented in the PCM \cite{Lang:23slp}. Finally, TLWP is strongly inhibited in our 1D experiment but 3D modeling may induce more complex situations, even though the latent heat cooling associated with even the faintest evaporation rates likely prevents the temperature of surface ice deposits from reaching 273~K \cite{Schorg:20}.


\section{Temperature-dependent water contact parameter}
\label{section:tempmsensi}

\subsection{Sensitivity of the water cycle to the contact parameter}
\label{subsec:sensim}

\begin{figure}[h!]
\hspace{-80pt}
\includegraphics[width=1.5\textwidth]{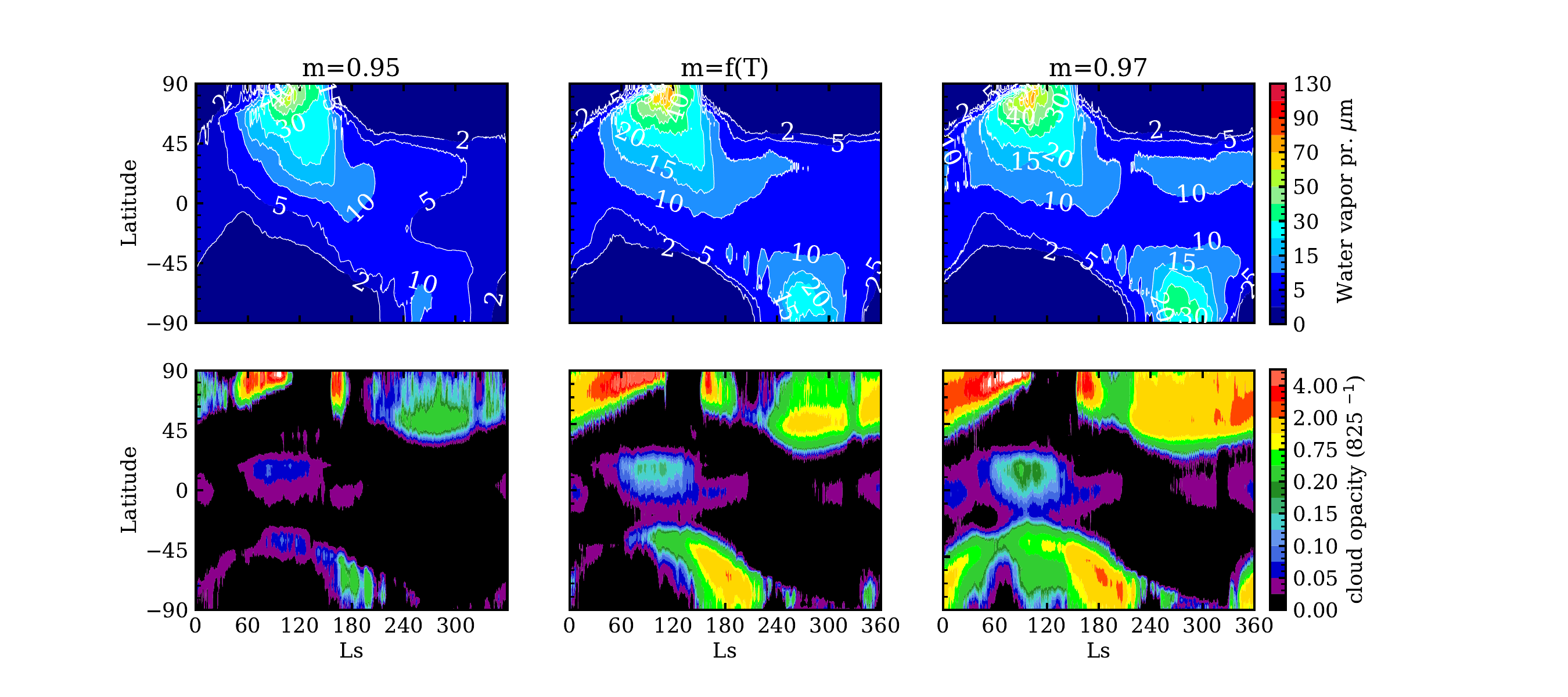}
\caption{Zonally averaged water vapor column (top) and cloud opacity (bottom) at 2 pm after a 10 year simulation for MY26 baseline water cycle parameters ($\nu_{\mbox {\scriptsize eff}}$=0.1, $\alpha$=0.35, TI=800~tiu, as in table \ref{table:watparams}) for a constant reference water contact parameter of 0.95 (left), a temperature-dependent contact parameter (middle) and a constant 0.97 contact parameter (right).}
\label{fig:sensim}
\end{figure}

The influence of the contact parameter on the global water cycle is caused by a local effect. For a given amount of water vapor, at a given temperature, the PCM computes the amount of forming ice (if any). Provided that not all the available nuclei are used, the contact parameter describes the fraction of ice which either accumulates on existing crystals or activates nuclei to form new particles. Increasing the efficiency of nucleation, which corresponds to an increase of the contact parameter in our scheme, favors formation of new particles over crystal growth. Water ice particles are thus smaller and more effectively transported outside the polar regions, effectively controlling the total amount of water in the atmosphere throughout the year. This is illustrated in side panels of figure \ref{fig:sensim}~: the baseline settings produce a dry water cycle (left panels). Increasing the contact parameter from 0.95 to 0.97 (right panels) allows for a better representation of the key features of the water cycle (northern sublimation peak, aphelion cloud belt opacity, northern fall tropical humidity) but also increases the opacity of clouds during the polar night, which is already a well-known caveat of the PCM.
This is extensively discussed in \citeA{Nava:14jgr}. 

We have tested all the proposed temperature-dependencies formulations from \citeA{Maat:14} shown in figure \ref{fig:temp_dep_m}, and we find that the linear relation reported in equation \ref{eq:tempdepm} allows for the best water cycle features without disrupting our entire water cycle tuning study (middle panels). This formulation indeed produces sustained cloud nucleation in the aphelion cloud belt atmospheric temperature range, while strongly diminishing both the northern and southern polar hoods and maintaining a reasonable global inventory of water in the atmosphere. As we are manually tuning each parameter, we have chosen this simple parameterization. As stated in \citeA{Maat:14}, the use of an hyperbolic relation for the temperature dependency is probably the most realist depiction of the behavior of cloud nucleation on Mars, which is dramatically inhibited at low temperature. Having a better representation of this in the PCM (especially in the polar night) would require tuning several water cycle parameters simultaneously. Since $\nu_{\mbox {\scriptsize eff}}$ is poorly constrained both in theory and with observations, and the water cycle in the PCM is so sensitive to any change in the tunable parameters, we have chosen to only investigate the contact parameter. Due to the sensitive retroactions with the radiatively active clouds, investigating the water cycle with coupled variations in all the model tunable parameters is too complex to be studied with a few simulations and calls for dedicated numerical methods.

\subsection{Ice particle size}
\label{subsec:mrsize}

The Mars PCM notoriously overestimates water ice particle size \cite{Nava:14jgr,Stch:22}. As explained in the previous section, the contact parameter affects the whole water cycle through water ice particle size variations. All other things being equal, the use of a temperature-dependent nucleation scheme is expected to have a direct local effect on ice particle size~: within warm clouds, e.g. in the Aphelion Cloud Belt, nucleation is favored over crystal growth, the number of ice particles is increased and ice particle size is reduced. Conversely, high latitude clouds have a reduced number of ice particles than with a constant contact parameter, and thus with increased radius. In addition, as the whole system is extremely sensitive to the presence of clouds, one could expect the new contact parameter scheme to affect the vertical distribution of clouds, or any atmospheric parameter which would in turn have an indirect effect on water ice particle size.

To assess the current capabilities of the PCM on this matter, we compare our outputs with ACS-MIR retrieved data for MY35 \cite{Stch:22}. MY35 has an atmospheric dust scenario comparable to MY26 after $L_S$=60$^\circ$\cite{Mont:15,Mont:20}~: we interpolate our simulations outputs of MY26 at the same location, local time and local altitude as the observations. Due to the orbital configuration, only 3 profiles are available within the Aphelion Cloud Belt. They show a decrease of around $\sim$20~\% in particle sizes simulated with our new temperature-dependent scheme. In these cases, the vertical profiles are retrieved between 20 and 50~km altitude with effective water ice radius up to $\sim$4~$\mu$m at 20~km. Both our new reference simulation and the previous settings by \citeA{Nava:14jgr} show water ice radius up to 8~$\mu$m at these altitudes, and going up to 50-60 $\mu$m respectively around 10~km.

\begin{figure}[!htb]
\begin{center} \includegraphics[width=0.9\textwidth]{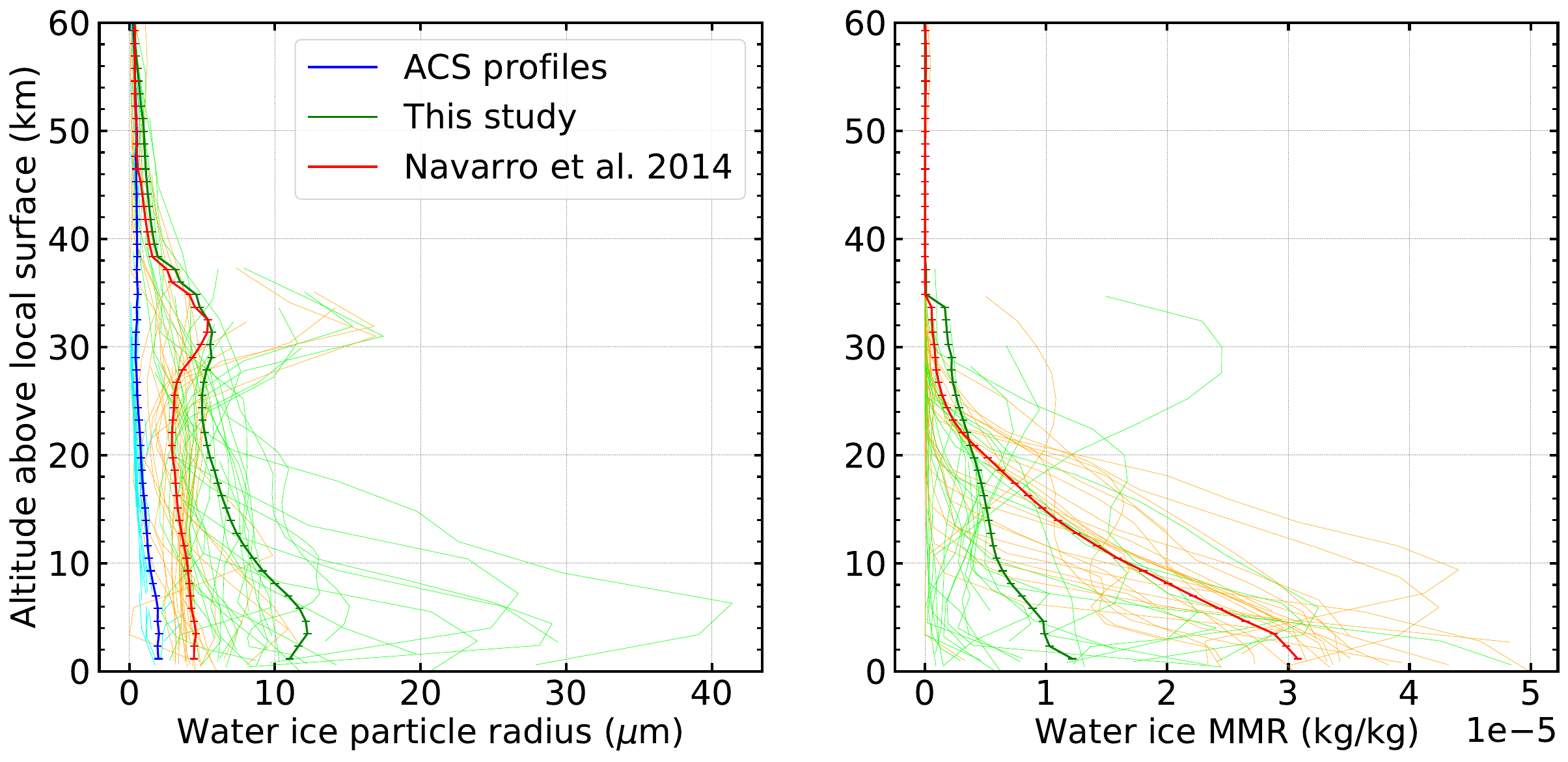}
\end{center}
\caption{ACS-MIR retrievals for vertical profiles of water particles effective radius in the 2.3-4.2 $\mu$m spectral range (in lightblue, mean profile in blue) at the Northern polar night edge between $L_S$ 190-240$^\circ$ of MY35 (left panel). Individual profiles at the same location and local time are interpolated from PCM outputs for our new reference simulation (lightgreen, mean profile in green) and simulation using the settings of \citeNP{Nava:14jgr} (orange, mean profile in red). The corresponding water ice mass mixing ratio are computed for both simulations (right panel).}
\label{fig:acs_rice}
\end{figure}

Besides, figure \ref{fig:acs_rice} shows a selection of profiles between $L_S$=190-240$^\circ$ in the Northern hemisphere, at latitudes over 50$^\circ$N corresponding to the edge of the polar hood, with associated mass mixing ratios. Column integrated cloud optical depth is similar in both simulations (around 1.5 at 825~cm$^{-1}$) in this region at this time of year, which translates in a reduced cloud mass but increased ice particle size with our new temperature-dependent contact parameter. Comparison between data and PCM output is not straightforward, as we investigate effective radius size versus radius size computed in the PCM, as a mean geometric radius. Furthermore, retrieved effective particle size may be underestimated at low altitude, at the end of the line of sight of the instrument. Nevertheless, the PCM still has a strong overestimating bias, especially at the bottom of the clouds. The physical reasons remain elusive at the moment. 

The discrepancy between retrievals and model outputs is also partly due to the difficult comparison between effective radius, which refers to the measured opacities \cite{Clan:03,Wolf:03,Stch:22}, and modeled ice particle size which is a local computation without considering any observational configuration. Even though some recent publications suggest that some water ice particle radii may be well over $\sim$15~microns \cite{Lemm:22,Wolf:22} in the Aphelion Cloud Belt, additional efforts in building comparable datasets between retrievals and models are needed to unravel the observed mismatch.

\section{Perennial ice and seasonal frost albedo discrepancy}
\label{section:albedo}

In order to refine our representation of the atmospheric water content, we introduce a difference between the albedos of perennial water ice and seasonal frost. In the PCM ``seasonal frost'' encompasses all types of atmospheric water ice surface deposition (direct surface condensation, sedimentation, turbulent mixing, or also scavenging by CO$_2$ ice (see section \ref{section:scavenging})). Perennial ice corresponds to infinite ground water ice reservoirs, which are spatially prescribed in the PCM.

If there is seasonal frost over perennial ice grid cells, it has to be entirely sublimated before underneath perennial ice can interact with the atmosphere. So far, the only difference between the two ice types is the albedo, and future improvements may include additional dichotomy regarding their physical properties, such as thermal inertia, surface ice coverage in the model mesh, etc.

Several processes may influence the evolution of frost albedo on Mars. Dust accumulation will darken surface frost, and the ice grain size distribution will also change over time and affect reflective properties of the ice, lowering its albedo. Thus, frost has a higher albedo when freshly deposited. We simply assume only two possible albedos, either for frost or perennial ice. 

\subsection{Determination of the frost metamorphism threshold}

\begin{figure}[!htb]
\hspace{-110pt}
\includegraphics[width=1.4\textwidth]{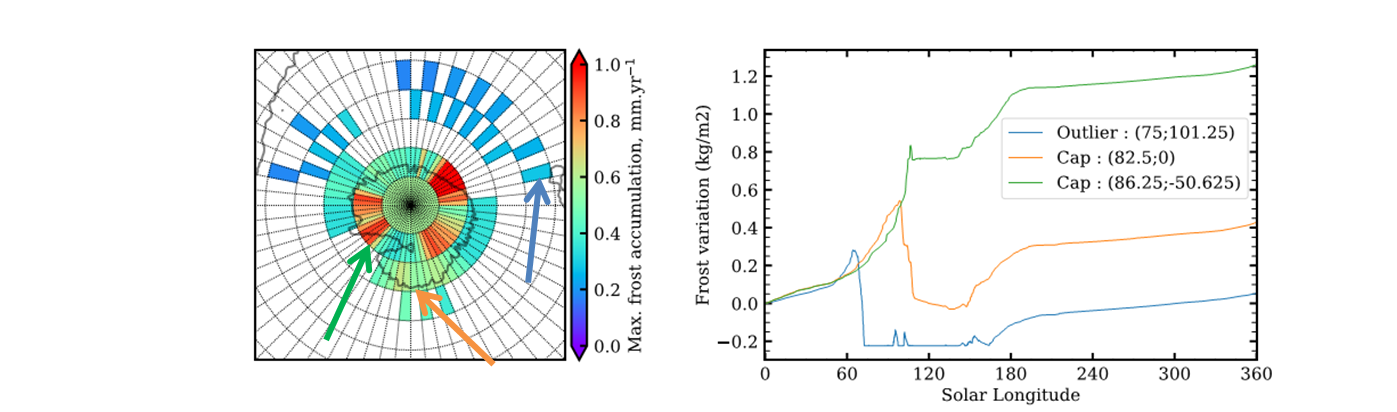}
\caption{\textbf{Left:} Maximal seasonal frost accumulation from the beginning of the year [mm, or kg.m$^{-2}$], on the Northern Polar Cap area in the PCM. Colored arrows refer to the local grid cells studied on the right panel. \textbf{Right:} Annual frost variations [kg/m²] from the start of the year. When the surface frost is depleted (around sol 150 for the outlier grid cell), perennial ice starts subliming with a higher albedo.}
\label{fig:regimes}
\end{figure}

In the PCM on the Northern Polar Cap, we identify three types of seasonal frost behavior over perennial ice, as shown in figure \ref{fig:regimes}~: continuous frost deposition (on the main cap); annual net frost depletion (mainly in the outlier region); and a first subliming phase during Northern Summer followed by frost accumulation leading to an annual net frost deposition (in the outer main cap). In the last two scenarios, the maximum amount of surface frost sublimated is around 0.5 kg.m$^{-2}$.

As explained in section \ref{subsub:alb} we introduce a ``frost metamorphism threshold'' to convert seasonal frost over the NPC into perennial ice. Since we observe a maximal seasonal frost deposition in the first half of the year of approximately 0.5~kg.m$^{-2}$ on subliming grid cells in figure \ref{fig:regimes}, we set the threshold to this value. Numerically this means that when frost accumulation exceeds the threshold, the surplus below is converted into perennial ice, as seen on figure \ref{fig:schemaalb}. This simple parameterization is more realistic than the previous PCM version and may be improved in the future by accounting for the dust deposition rate or the evolution of grain size with temperature, with a time-varying albedo for frost.\newline

\subsection{Tuning the Northern sublimation peak with differentiated albedos}

\begin{figure}[h!]
\includegraphics[width=\textwidth]{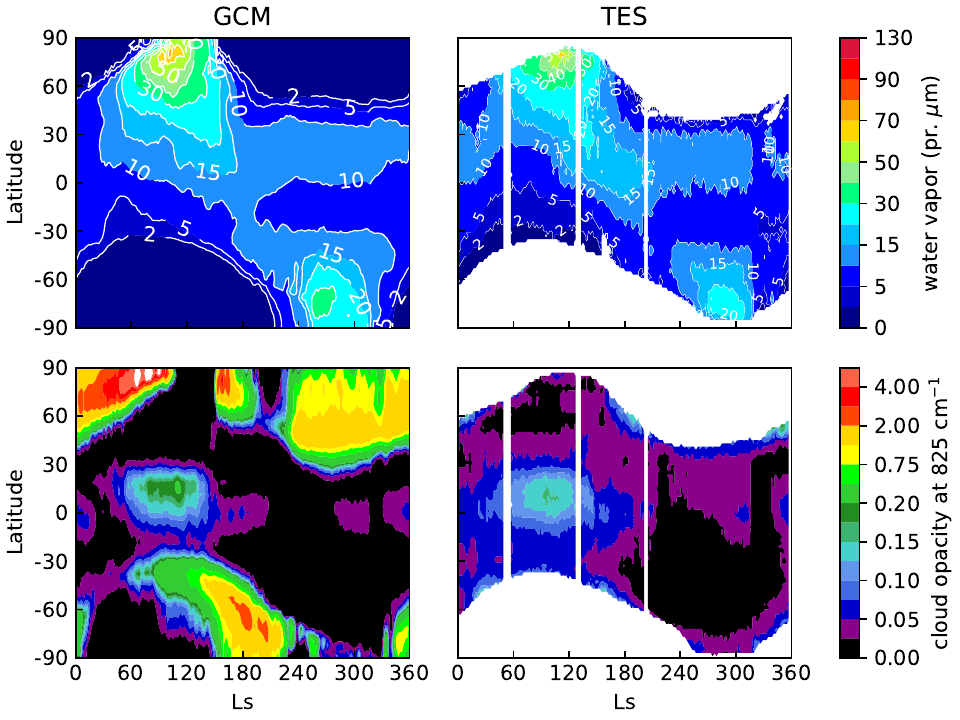}
\caption{Zonally averaged quantities at 2 pm local time for water vapor column in the PCM (top left) and corresponding TES observations (top right), and cloud opacity in the PCM (bottom left) with corresponding TES observations (bottom right). Results are shown after 10 years of simulation with a frost albedo of 0.35 and a perennial ice albedo of 0.32, a frost metamorphism threshold of 0.5 kg.m$^{-2}$, and initial conditions corresponding to figure \ref{fig:al033_wc} reference steady-state simulation.}
\label{fig:ap032f035}
\end{figure}

Using this parameterization with a fixed frost metamorphism threshold, we have tested several set of albedos for perennial ice and frost albedo above the NPC. Figure \ref{fig:ap032f035} shows the best correspondance we have with TES observations for integrated water vapor column and cloud opacity, with a frost albedo of 0.35 and a perennial ice albedo of 0.32.

During summer sublimation over the NPC, a fraction of the water vapor immediately condenses back on the coldest part of the cap. The mere use of two albedos manages to reduce the overall intensity of the sublimation event and produce an integrated water vapor column comparable to TES observations. The seasonal evolution of albedo over the NPC is much more complex, with different regional behaviors \cite{Hale:05}, than the one modeled in the PCM with our parameterization. However, since our present model configuration can not properly describe the details of the water cycle over the NPC \cite{Pott:17}, our metric is the sublimation peak around $L_S$=120$^\circ$, which is strongly improved compared to our reference simulation.

\section{Scavenging by CO$_2$ ice}
\label{section:scavenging}

In this section, we explore the impact of the scavenging of water ice and dust particles by CO$_2$ atmospheric condensation and fall in the polar nights on the global water cycle. The parametrization we use here is simpler but less numerically expensive than the full microphysical CO$_2$ cloud model that has been developed for the PCM by \citeA{Maat:22}.

\subsection{Determination of the scavenging ratio}

As detailed in section \ref{subsubsec:param_scavenging}, in the model, the scavenging of water ice particles by the CO$_2$ snow relies on one coefficient, the scavenging ratio. On Earth, several measurement campaigns have been carried out to estimate its variations for hydrometeors, and yield a wide range of values that depend on the aerosol and type of precipitation. For instance, \citeA{Jaff:90} measured ${\cal R}_{{\mbox {\scriptsize scav}}}$ varying between 500 and 1500 in liquid water clouds near Paris, France. \citeA{Gogo:18} also reported values around 120 in Antarctic water ice clouds. In our Martian PCM, we leave it as a tunable parameter and explore the climate sensitivity to its value for the CO$_2$ snow. Due the prevailing heterogeneous mode for CO$_2$ nucleation \cite{Maat:10}, the enhanced radiative cooling that occurs in the vicinity of the aerosols in the night, or also the scavenging induced by collisions during the fall, the ratio ${\cal R}_{{\mbox {\scriptsize scav}}}$ is more than likely to be higher than 1. However, there is no direct observational constraints on this scavenging ratio for the CO$_2$ snow on Mars, although some recent studies like \citeA{Alsa:22} give some elements for indirect validation.

We thus tested values for ${\cal R}_{{\mbox {\scriptsize scav}}}$ of 10, 20, 50 and 100. We assign the same value of ${\cal R}_{{\mbox {\scriptsize scav}}}$ for both dust and water ice aerosols. This simplification limits the number of free parameters in the model. We also conducted a simulation where we activate the scavenging by CO$_2$ only on water ice particles, with ${\cal R}_{{\mbox {\scriptsize scav}}}$=20.

\begin{figure}[!htb]
\begin{center} \includegraphics[width=\textwidth]{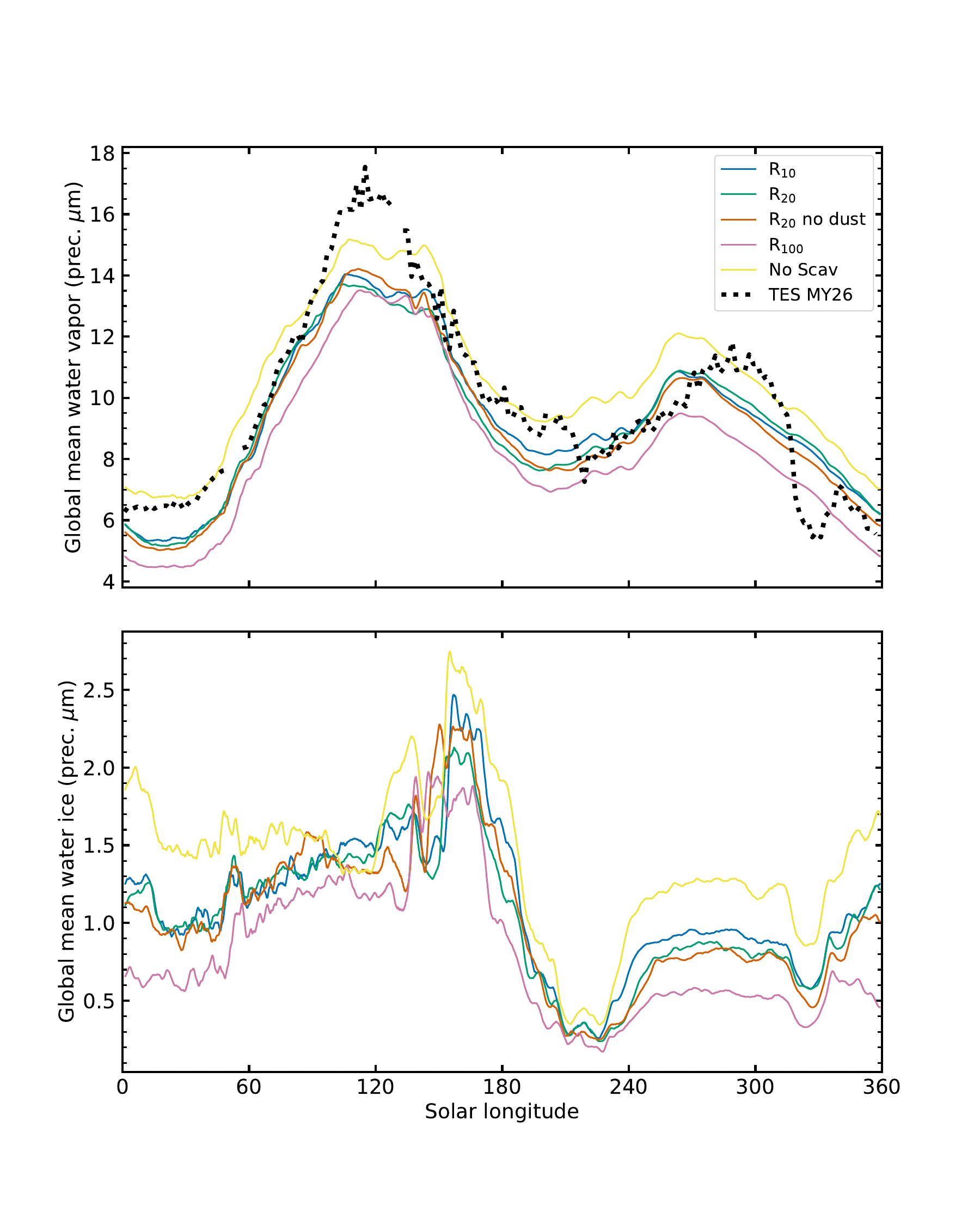}
\end{center}
\caption{Globally-averaged atmospheric water vapor (top panel) and ice (bottom panel) for the investigated scavenging ratios of 10 (blue), 20 (green), 20 for water ice only (red), and 100 (purple). The simulation with ${\cal R}_{{\mbox {\scriptsize scav}}}$=50 is not shown. Reference curve (yellow) is the simulation presented in section \ref{section:refsimu}, without the scavenging by CO$_2$.}
\label{fig:scavratio_gwc}
\end{figure}

Figure \ref{fig:scavratio_gwc} presents the corresponding planetary mean atmospheric water vapor and water ice columns for the different values of ${\cal R}_{{\mbox {\scriptsize scav}}}$, after 10 cycled MY26 simulations. Indeed, the scavenging of the water ice clouds by the CO$_2$ snow, especially in the Northern polar night, triggers the radiative water ice cloud feedback, already mentioned in this article. The removal of clouds in the polar hood ends up drying the whole water cycle, which takes several years to stabilize toward a new equilibrium. The simulations with ${\cal R}_{{\mbox {\scriptsize scav}}}$ 10 and 20 quickly converge to an equilibrium level of about 20\% dryer than the reference water cycle at $L_S$ 0$^\circ$, although this comparison slightly varies with season. On the other hand, a ${\cal R}_{{\mbox {\scriptsize scav}}}$ of 100 steadily dries up the model and is not fully stabilized even after 10 years of simulation, reaching -38\% at the beginning and -40\% at the end of the 10th year. The simulation with ${\cal R}_{{\mbox {\scriptsize scav}}}$=50 (not shown) exhibits a similar behaviour.

\subsection{Influence of the scavenging on the water cycle}

\begin{figure}[!htb]
\begin{center} \includegraphics[width=\textwidth]{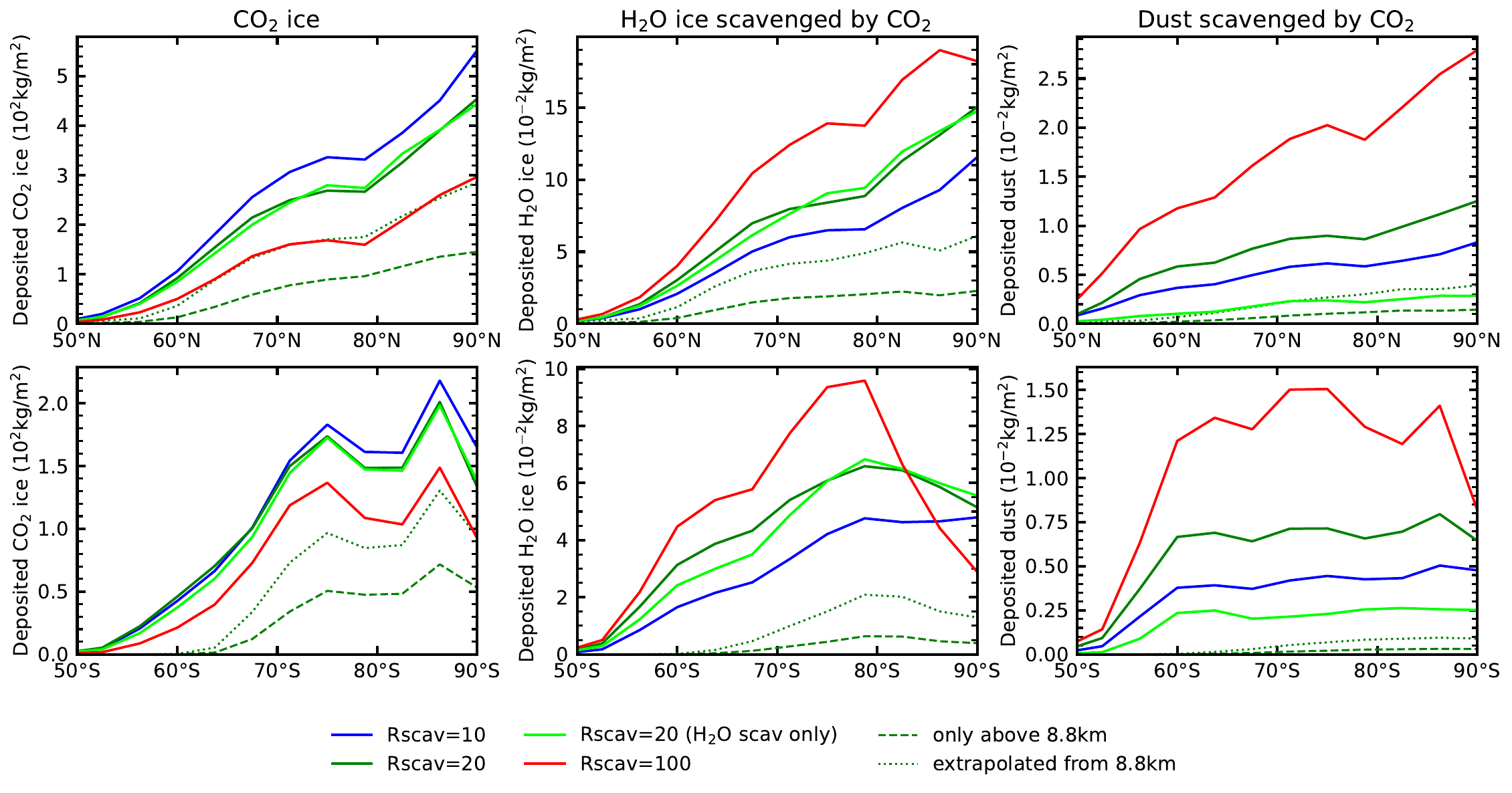}
\end{center}
\caption{Total CO$_2$ ice fall and aerosol scavenging per latitude (zonal mean, annual accumulation) during MY26 simulations, respectively at the North (top) and South (bottom) poles. Solid lines represent the mass that is deposited onto the surface from the whole atmospheric column, while dashed lines only account for the mass falling from above an altitude of 8.8km (plotted only for ${\cal R}_{{\mbox {\scriptsize scav}}}$=20). The dotted lines represent what would be deposited from the whole column when extrapolating the mass mixing ratio tendency due to forced falling (for CO$_2$ ice) or scavenging (for H$_2$O ice and dust) from the 8.8km level down to the surface.}
\label{fig:qscav_comp}
\end{figure}

By definition, for a given quantity of condensing CO$_2$, and aerosol mixing ratio in the cell, the amount of scavenged aerosol varies accordingly to the chosen scavenging ratio. However, this scavenged aerosol amount is not striclty proportional to the scavenging ratio, and a simulation with a twice higher ${\cal R}_{{\mbox {\scriptsize scav}}}$ does not double the aerosol scavenging by CO$_2$ snow (figure \ref{fig:qscav_comp} and table \ref{tab:qscav_comp}). This suggests that some retroactions compensate for the loss of aerosols in the atmosphere. One example appears clearly on the left panels of figure \ref{fig:qscav_comp}~: when H$_2$O ice and dust are more scavenged (bigger ${\cal R}_{{\mbox {\scriptsize scav}}}$), there is less CO$_2$ condensation. In fact, the radiative cooling in the polar night is strongly enhanced by the presence of H$_2$O ice and dust aerosols \cite<e.g.>{Forg:98}, making the CO$_2$ condense preferentially in highly aerosol-loaded regions. The scavenging thus acts as a negative feedback for the CO$_2$ condensation.

As shown on figure \ref{fig:qscav_comp}, the CO$_2$ condenses more as we get closer to the poles, with a general increase with latitude in the North up to the pole (with a slight local maximum at 75$^\circ$N), while the South seems to reach a first peak near 75$^\circ$S and a second of similar amplitude at 86$^\circ$S. In absolute mass (kg), this is of course different, as the greater surface covered by the lower latitudes compared to the pole makes the CO$_2$ total mass deposition peak around latitudes of $\pm$70$^\circ$.
The water ice deposition due to scavenging follows the meridional trend of CO$_2$ ice in the North, and to a lesser extent in the South for low scavenging ratios ${\cal R}_{{\mbox {\scriptsize scav}}}$=10, 20. For ${\cal R}_{{\mbox {\scriptsize scav}}}$=100, the H$_2$O ice deposition peaks just below 80$^\circ$S, before drastically dropping closer to the pole. Further analysis shows that water ice clouds hardly reach these high latitudes in this simulation, as they are fainter (figure \ref{fig:scavratio_gwc}) and are probably scavenged before being transported up there.
For the low ${\cal R}_{{\mbox {\scriptsize scav}}}$, dust is scavenged $\sim$8 times less than water ice. Among this quantity, the comparison of the simulations with and without the scavenging of dust (figure \ref{fig:qscav_comp}, right panels), shows that between 1/4 and 1/3 of the dust is actually being deposited as water ice cloud nuclei (trapped in the water ice particles, which in turn, are being scavenged by the CO$_2$ snow). This difference between the water and dust deposition actually corresponds quite well to the ratio of their respective local atmospheric concentrations where CO$_2$ condensates. Indeed, the dust scenario, to which the PCM dust column is renormalized \cite{Made:11dust}, features very little dust amount in the polar nights \cite{Mont:15}, while water ice clouds form an opaque polar hood. Activating or not the scavenging on the dust particles also does not affect much the other diagnoses (figure \ref{fig:scavratio_gwc}, and left and middle panels of figure \ref{fig:qscav_comp}, ${\cal R}_{{\mbox {\scriptsize scav}}}$=20 simulations). 
This would potentially differ in the case of greater dust loading, or lesser water ice clouds (as is the case for the ${\cal R}_{{\mbox {\scriptsize scav}}}$=100 simulation, where the ratio of scavenged H$_2$O over scavenged dust decreases to $\sim$5 overall, see table \ref{tab:qscav_comp}).

Figure \ref{fig:qscav_comp} also provides insights on the vertical distribution of the CO$_2$ atmospheric condensation and aerosol scavenging, specifically in the ${\cal R}_{{\mbox {\scriptsize scav}}}$=20 simulation. The CO$_2$ polar clouds can feature a diversity of vertical structures, depending for instance on the surface winds and wave activity \cite{Tobi:03}. They have been observed \cite{Alsa:22} and simulated \cite{Maat:22} to reach up to 40km in the polar atmosphere. In our simulations, some CO$_2$ condenses up to this altitude as well, but the main condensation levels and sources for CO$_2$ ice deposition at the surface lie below 10km. At both poles, the extrapolated condensation tendency from above 8.8km down to the surface can only make up to half of the total column CO$_2$ deposition. This is even less in regard to the aerosol scavenging, which also depends on their own vertical distribution~: in the North, the extrapolated scavenging tendency from above 8.8km would account for less than half of the total deposition for H$_2$O and 1/3 for dust ; in the South, it would drop to 1/4 for H$_2$O and 1/10 for dust.

\subsection{Budget of scavenged water ice}

The work from \citeA{Alsa:22} provides an interesting element of comparison for our parametrization. They derived CO$_2$ ice opacity from measurements by Mars Reconnaissance Orbiter Mars Climate Sounder (MCS), and used a CO$_2$ snowfall model, with assumptions on the H$_2$O ice serving as nuclei for the CO$_2$ clouds. This allowed them to estimate the budget of CO$_2$ and water ices deposited onto the surface via these processes. 
For CO$_2$, they evaluate the deposited snowfall to 3.4-4.6$\times$10$^{13}$ kg in the North and 0.34-2.1$\times$10$^{13}$ kg in the South. This is two orders of magnitude below our model outputs (see table \ref{tab:qscav_comp}), while their ratio between both poles ranges from similar to us up to one order of magnitude above, depending on the used values.
These differences are conveyed into the H$_2$O scavenging estimations, which they compute to be of 3.2-30$\times$10$^{9}$ kg in the North and 0.42-8.3$\times$10$^{9}$ kg in the South, about two orders of magnitude below the PCM predictions as well. The North-South asymmetry is also preserved, with a ratio ranging from 3.6 to 7.7, above those from our model.
Some elements can partially explain these differences~: for example, the gaps in MCS observations at lower altitudes calls for extrapolated CO$_2$ ice vertical profile. These extrapolations introduce an underestimating  bias regarding the amount of CO$_2$ ice deposition, according to our PCM results. This is even more true for the scavenged water ice, especially in the south, which therefore also increases the observed asymmetry between the poles. Conversely, the coverage used by \citeA{Alsa:22} is more limited than our analyses with the PCM~: spatially, MRO's orbit makes it difficult to run their analysis below 65$^\circ$N/S, while it seems CO$_2$ can condense at lower latitudes, which cover a greater area than the polar ones ; temporally, the analysis is also restricted to the winter hemisphere (i.e., $L_S$ 0$^\circ$-180$^\circ$ for the South and 180$^\circ$-360$^\circ$ for the North, respectively), whereas the PCM still predicts significant northern CO$_2$ snow fall and water ice scavenging between $L_S$ 0$^\circ$ and 40$^\circ$.
Eventually, \citeA{Alsa:22} state that a fraction of 1 to 15\% of the total Northern seasonal water ice deposits is due to the scavenging by CO$_2$ snowfall. With the PCM, which predicts $\sim$2$\times$10$^{12}$ kg (1.2$\times$10$^{12}$ kg) of total water ice surface frost deposition in the North (South) within the year, this percentage ranges from 27 to 57\% (29 to 74\%) depending on the scavenging ratio (with the highest fractions corresponding to ${\cal R}_{{\mbox {\scriptsize scav}}}$=100).

\begin{table}
\caption{Total CO$_2$ ice fall and aerosol scavenging due to CO$_2$ snow during MY26 simulations.}
\label{tab:qscav_comp}
\begin{tabular}{lcccccc}
\toprule
\multirow{2}{*}{PCM run}
  & \multicolumn{2}{c}{{\small CO$_2$ mass deposited (10$^{15}$kg)}} & \multicolumn{2}{c}{{\small H$_2$O mass deposited (10$^{11}$kg)}} & \multicolumn{2}{c}{{\small Dust mass deposited (10$^{11}$kg)}} \\
\cmidrule(r){2-3}
\cmidrule(r){4-5}
\cmidrule(r){6-7}
 & North & South & North & South & North & South \\
\midrule
{\small ${\cal R}_{{\mbox {\scriptsize scav}}}$=10}
    & 2.89 & 1.28 & 5.71 & 3.51 & 0.68 & 0.52 \\
{\small ${\cal R}_{{\mbox {\scriptsize scav}}}$=20}
    & 2.39 & 1.26 & 7.78 & 5.70 & 1.03 & 0.88 \\
{\small ${\cal R}_{{\mbox {\scriptsize scav}}}$=20 (H$_2$O only)}
    & 2.35 & 1.17 & 7.50 & 5.00 & 0.23 & 0.28 \\
{\small ${\cal R}_{{\mbox {\scriptsize scav}}}$=100}
    & 1.47 & 0.86 & 11.64 & 7.81 & 2.22 & 1.69 \\
\bottomrule
\end{tabular}
\end{table}

Besides the already mentioned considerations on the ${\cal R}_{{\mbox {\scriptsize scav}}}$ coefficient, which is supposed constant everywhere and identical for both dust and water ice, the parametrization in itself is quite simplified compared to a microphysics scheme that would represent CO$_2$ ice clouds, like the one from \citeA{Maat:22}. Specifically, the forced falling of CO$_2$ snow within a physical timestep span (7.5 minutes), from sometimes up to 40km, is very unrealistic. 
Horizontal transport and sedimentation kinetics that could allow CO$_2$ clouds to meet sublimating conditions, or could at least modify the meridional distribution of the surface deposition (which is currently somewhat different than inferred by \citeA{Alsa:22}) are neglected. The influence of a CO$_2$ ice cloud on local convection \cite{Cola:08} or the radiative transfer in the polar night \cite{Cola:03b} is also not represented. It is thus likely that the amount of water ice that is deposited at the surface by this parametrization is overestimated and constitutes an upper boundary for the CO$_2$-scavenged water ice budget. This process has nonetheless an indisputable effect on the global water cycle due to the RAC feedback, and motivates its representation in martian Global Climate Models, either via simple parametrizations like this one or more complete microphysical models \cite{Maat:22}.

\section{Conclusion and prospects}
\label{section:ccl}

We have implemented four new independent physical parameterizations~: \newline
 - the latent heat of ground ice sublimation. We find that, as expected, the associated energy flux over ice sublimation is in most cases negligible in our PCM, because of the low temperatures of perennial ice deposits in the polar regions and low amount of water frost elsewhere on the planet. However, this process is relevant for paleoclimatic studies where surface ice deposits may be tens of Kelvins warmer than the present-day Northern Polar Cap. Our study shows that, in the recent past of Mars, the formation of liquid water through direct melting of surface water ice at lower latitudes is inhibited by latent heat cooling. \newline
 - a temperature-dependent contact parameter for cloud nucleation in our microphysics scheme. This contact parameter strongly diminishes the opacity of water ice clouds within the polar hood, which has been a well known defect of the PCM, while maintaining a thick-enough Aphelion Cloud Belt. The could-free season in the Northern polar night observed by PFS and suggested in MCS data is also predicted with the inhibition of low-temperature nucleation. This realistic behavior of nucleation is an improvement regarding the representation of supersaturated water vapor but fails to address the overestimation of water ice particle size, especially at the bottom of clouds. The origin of the bias in water ice particle size in the PCM remains elusive for now.\newline
 - a separation between frost and perennial ice, restricted to their albedo at the moment. Because of the coarse representation of the Northern Polar Cap, its seasonal behavior in the PCM does not match the observations. Tackling this issue probably requires accounting for local variations of thermophysical properties of the ice reservoirs atop the Northern polar cap, which is out of the scope of this study. Nevertheless, we show that this parameterization dampens the amplitude of the Northern sublimation peak to the observed TES values.\newline
 - the scavenging of atmospheric water ice by CO$_2$ ice formation and falling in the polar night. Parameterizing this process has a huge indirect impact on the global water cycle, due to the radiative polar clouds feedbacks. 
 For the dust cycle, the column opacity is normalized to a daily target scenario that features very little dust in the polar night. The scavenging by CO$_2$ snow thus does not impact much its already low concentration. However, with the latest release of a more self-consistent dust cycle in the PCM (publication in preparation), parametrizing this process becomes all the more necessary, as it is probably one of the causes for such a low dust opacity in the observational scenario in the first place.\newline

The development of the PCM historically consists in incremental improvements of the physics. The philosophy is to implement realistic processes with relatively simple schemes, with a low number of tunable parameters. Doing so, we do aim less for a perfect agreement with observations than for good physical comprehension of the behavior of our model. Parameters are tuned manually and independently. As illustrated in this paper, the water, dust and CO$_2$ components of the climate system have non-linear interactions and retroactions with one another. Any change in atmospheric water inventory, resulting from direct or indirect effect of related parameterizations, is reflected in cloud distribution which is in turn subject to radiatively active cloud feedback. Diagnosing the interactions at work is becoming increasingly complex as the number of represented physical processes grows~: numerical modeling of the martian climate is becoming a science field in itself. We feel like that our traditional approach of manually tuning our model could strongly benefit from a complementary approach inspired by terrestrial climate modelers with machine learning tuning tools \cite<e.g.,>{Couv:21,Hour:21,Vill:21}, to try and overcome the limitation of parameters being tuned independently and render how they interact. This is especially true for the water cycle, which is extremely sensitive to any input because of the strong radiatively active cloud feedbacks. The study of the global water cycle on Mars could thus be investigated, like in Earth communities, as an emerging property of all the individual components of the system \cite{Brie:20}.

\section*{Open Research Section}

ACS data used to evaluate modeled ice particle size can be found at \url{https://data.mendeley.com/datasets/hff8kc6fsy/1} \cite{Stch:22dataset}. The numerical simulations were conducted with the Planetary Climate Model (previously LMD-GCM), developed in Laboratoire de Météorologie Dynamique and available to the public at \url{https://web.lmd.jussieu.fr/~lmdz/planets/mars/}. All the simulations were performed using the r2840 version of the model. The model outputs and analysis scripts we wrote are archived at Zenodo \cite{Naar:23dataset1} and publicly available.

\acknowledgments
This project has received funding from the European Research Council (ERC) under the European Union’s Horizon 2020 research and innovation program (grant agreement No 835275, project ``Mars Through Time'').  Mars PCM simulations were done thanks to the High-Performance Computing (HPC) resources of Centre Informatique National de l’Enseignement Supérieur (CINES) under the allocation n\textdegree A0100110391 made by Grand Equipement National de Calcul Intensif (GENCI).


%
%

\bibliography{bib_joseph}

%
%
%
%
%

\end{document}